\documentclass[aps,prb,reprint,superscriptaddress,amsmath,amssymb]{revtex4-2}
\usepackage{graphicx}
\usepackage{dcolumn}
\usepackage{bm}

\begin{document}


\title{Distorted triangular skyrmion lattice in a noncentrosymmetric tetragonal magnet}

\author{Takeshi Matsumura}
\email[]{tmatsu@hiroshima-u.ac.jp}
\affiliation{Department of Quantum Matter, ADSE, Hiroshima University, Higashi-Hiroshima 739-8530, Japan}
\author{Kenshin Kurauchi}
\affiliation{Department of Quantum Matter, ADSE, Hiroshima University, Higashi-Hiroshima 739-8530, Japan}
\author{Mitsuru Tsukagoshi}
\affiliation{Department of Quantum Matter, ADSE, Hiroshima University, Higashi-Hiroshima 739-8530, Japan}
\author{Nonoka Higa}
\affiliation{Department of Quantum Matter, ADSE, Hiroshima University, Higashi-Hiroshima 739-8530, Japan}
\author{Hironori Nakao}
\affiliation{Photon Factory, Institute of Materials Structure Science, High Energy Accelerator Research Organization, Tsukuba, 305-0801, Japan}
\author{Masashi Kakihana}
\affiliation{Faculty of Science, University of the Ryukyus, Nishihara, Okinawa 903-0213, Japan}
\author{Masato Hedo}
\affiliation{Faculty of Science, University of the Ryukyus, Nishihara, Okinawa 903-0213, Japan}
\author{Takao Nakama}
\affiliation{Faculty of Science, University of the Ryukyus, Nishihara, Okinawa 903-0213, Japan}
\author{Yoshichika \={O}nuki}
\affiliation{Faculty of Science, University of the Ryukyus, Nishihara, Okinawa 903-0213, Japan}
\affiliation{RIKEN Center for Emergent Matter Science, Wako, Saitama 351-0198, Japan}


\date{\today}

\begin{abstract}
Magnetic skyrmions are particle-like spin-swirling objects ubiquitously realized in magnets. 
They are topologically stable chiral kinks composed of multiple modulation waves of spiral spin structures, where the helicity of each spiral is usually selected by antisymmetric exchange interactions in noncentrosymmetric crystals. 
We report an experimental observation of a distorted triangular lattice of skyrmions in the polar tetragonal magnet EuNiGe$_3$, reflecting a strong coupling with the lattice. 
Moreover, through resonant x-ray diffraction, we find that the magnetic helicity of the original spiral at zero field is reversed when the skyrmion lattice is formed in a magnetic field. 
This means that the energy gain provided by the skyrmion lattice formation is larger than the antisymmetric exchange interaction. 
Our findings will lead us to a further understanding of emergent magnetic states. 
\end{abstract}

\maketitle

\section{Introduction}
Magnetic moments in crystals prefer to align themselves in various kinds of self-organized structure to minimize the total free energy at finite temperatures. 
Among these structures, magnetic skyrmion lattice, a periodic arrangement of particle-like spin-swirling objects realized in magnetic fields, is of special interest because of its nontrivial and spectacular structure~\cite{Nagaosa13,Tokura21}. 
Since the first discovery and demonstration of triangular skyrmion lattice (SkL) formation in MnSi~\cite{Muhlbauer09}, many types of SkLs have been reported.  Initially, they were discovered in chiral magnets without either a space inversion or a mirror reflection symmetry. 
In metallic B$_{20}$-type compounds with the $P2_{1}3$ space group, such as MnSi, Fe$_{1-x}$Co$_x$Si~\cite{Yu10,Munzer10}, and FeGe~\cite{Yamasaki15,McGrouther16}, with helical magnetic orderings at zero field, the Dzyaloshinskii-Moriya (DM)-type antisymmetric  interaction (ASI) in the form of $\bm{D}\cdot(\bm{S}_i \times \bm{S}_j)$, which prefers to twist the spin alignments, was considered an important ingredient for the formation of such unusual ordered states. 
The range of the SkL formation extends to insulators such as Cu$_2$OSeO$_3$~\cite{Seki12,Adams12}. 
The N\'{e}el type SkL, originating from the cycloidal nature of the spiral, was found in nonchiral polar crystals such as GaV$_4$S$_8$~\cite{Kezsmarki15} and VOSe$_2$O$_5$~\cite{Kurumaji17,Kurumaji21}, as has been theoretically predicted~\cite{Bogdanov89,Bogdanov94}. 
These SkL states generally have a much longer period than the lattice constant, resulting in effective decoupling of the magnetic and crystal structures. 
The propagation vectors of the constituent waves are almost always perpendicular to the external magnetic field. 

Recently, short-period SkL states have been discovered in rare-earth compounds. For example, in cubic EuPtSi with the same space group $P2_{1}3$ as MnSi, a very similar triangular SkL as that of MnSi is observed with a much extended phase stability down to very low temperatures~\cite{Kakihana18,Kaneko19,Tabata19}. 
As in MnSi, the magnetic order at zero field is helical, which is described by a single wave-vector ($\bm{q}$) and a single helicity. 
The DM-type ASI is therefore considered to play an essential role. Here, since the $S=7/2$ spin of the $4f$ electrons of Eu$^{2+}$ is well localized, the magnetic exchange interaction is mediated by the conduction electrons, which is called the Ruderman-Kittel-Kasuya-Yosida (RKKY) interaction~\cite{Hayami17}. 
Furthermore, recent discoveries of triangular SkLs in centrosymmetric rare-earth compounds such as Gd$_2$PdSi$_3$~\cite{Kurumaji19} and Gd$_3$Ru$_4$Al$_{12}$~\cite{Hirschberger19} with hexagonal lattices further extended the research field. 
Since the ASI is not expected in these compounds, magnetic frustration is suggested to be a possible origin of SkL formation~\cite{Okubo12,Leonov15}. 
A triangular SkL with triple-$\bm{q}$ constituent waves is compatible with the sixfold or threefold symmetry of the lattice. 
SkL states are observed even in tetragonal centrosymmetric compounds. 
In GdRu$_2$Al$_2$ and EuAl$_4$, a square and a rhombic SkL, respectively, are realized~\cite{Khanh20,Khanh22,Takagi22}, both indicating a strong coupling with the underlying crystal lattice. 
Theoretically, the higher-order exchange term of the RKKY interaction, the biquadratic term, is considered to be responsible for the multi-$\bm{q}$ magnetic structure and stabilizes the SkL state~\cite{Hayami21,Yambe22}. 

In contrast to the symmetric arrangements of the abovementioned SkLs, we report here the observation of a distorted triangular SkL realized in the tetragonal magnet EuNiGe$_3$ belonging to the polar space group $I4mm$, where the DM-type ASI is expected. 
The crystal structure is shown in Fig. 1A. 
The distortion of the triangle is a consequence of strong coupling between the magnetic structure and the tetragonal crystal lattice. 
Since the SkL generally prefers to form a triangular lattice to minimize the total free energy, the deformation shows that the spin system spontaneously chooses asymmetry to minimize the free energy in the tetragonal space. 
Moreover, we also show that the magnetic helicity of the original helimagnetic structure at zero field is uniquely determined in each magnetic domain and perfectly reflects the symmetry of the crystal. 
This shows that the DM-type ASI exists and fixes the helicity at zero field.
Then, we show that when the triangular SkL is formed in magnetic fields applied along the fourfold $c$-axis, all three constituent waves of the helimagnetic structure are unified to have the same helicity, in which one of the helicities is reversed from the zero-field helicity to match the primary wave. 
These results show that the energy gain to form the triangular SkL is larger than the DM-type ASI to twist the spins, suggesting that the latter is not the main driving force for SkL formation.

The magnetic properties of EuNiGe$_3$ have been well studied and are summarized in Fig. 1~\cite{Maurya14,Fabreges16,Kakihana17,Iha20}. 
A magnetic phase transition to a helimagnetic order with a propagation vector $\bm{q}\simeq (0.25, 0.05, 0)$ occurs at $T_{\text{N}}=13$ K~\cite{Fabreges16}. 
The more precisely determined magnetic structure in our study is shown in Fig. 1C. The amplitude along the $c$-axis is approximately 1.9 times larger than those in the $ab$ plane, indicating weak easy-axis anisotropy along the $c$-axis. 
Furthermore, as shown later, the helical plane is not perpendicular to $\bm{q}$. 
The magnetization curve for $H \parallel c$ exhibits discontinuous transitions at 2 T and 2.9 T, followed by a continuous increase to the fully polarized ferromagnetic state above 3.9 T (top panel of Fig. 1D). The magnetic phase diagram for $H \parallel c$ is shown in Fig. 1B~\cite{Kakihana17}. 
An attractive feature is that in intermediate phase II between 2 T and 2.9 T, an abrupt increase in the resistivity and a decrease in the Hall resistivity appear (bottom panel of Fig. 1D). 
This reminds us of the appearance of certain magnetic structures with an emergent magnetic field caused by topologically stabel SkL-like structures. 
Revealing the detailed magnetic structure of this phase is the main purpose of our study.

We employed resonant x-ray diffraction (RXD) to observe the magnetic scattering from the ordered structure (Fig. 2A). 
By tuning the x-ray energy at an absorption edge of the target element, the Eu $L_2$-edge here, the scattering intensity from the magnetic order is enhanced, which enables us to detect normally weak magnetic scattering of photons. 
Moreover, the high spatial resolution of RXD using a synchrotron x-ray beam allows us to resolve the small difference and variation in the $\bm{q}$ vectors. 
By using a phase retarder system, we can also manipulate the incident x-ray polarization, which is linear in the horizontal scattering plane, to right-handed circular polarization (RCP) or left-handed circular polarization (LCP). By analyzing the variation in the magnetic scattering intensity as a function of the incident polarization state, we investigated the magnetic helicity of the spiral orderings.

\section{helical magnetic order at zero field}
At zero field, we confirmed that eight magnetic Bragg peaks are observed at $\bm{q}=(\pm \delta_1, \pm \delta_2, 0)$ around the fundamental lattice peak, reflecting the formation of four magnetic domains, as demonstrated by neutron diffraction~\cite{Fabreges16}.  
The domains are labeled A, B, C, and D, as shown in Fig. 2C. 
From more high-resolution measurements of the peak position by RXD, we obtained $\delta_1=0.26$ and $\delta_2=0.052$, which are indeed incommensurate. A typical rocking scan of the resonant Bragg peak is shown in Fig. 2D, exhibiting a sharp width of $0.08^{\circ}$. 
In our RXD study, we performed polarization analysis of the diffracted x-rays. We also performed a phase-retarder scan of the incident beam, thereby deducing the Fourier component of the magnetic propagation vector~\cite{SM}. 
The single-$\bm{q}$ helical magnetic structure at zero field is expressed as
\begin{equation}
\bm{M}(\bm{r}) = \bm{m}(\bm{q}) \exp (i\bm{q}\cdot \bm{r}) + \bm{m}(\bm{q})^{*} \exp (-i\bm{q} \cdot \bm{r}) 
\label{eq:1}
\end{equation}
where $\bm{m}_{\bm{q}}$ represents the Fourier amplitude. 
Typical results for the phase-retarder scan are shown in Fig. 2B, where the incident polarization is varied from horizontal linear polarization ($\pi$) to LCP, vertical linear polarization ($\sigma$), RCP, and $\pi$ by rotating the angle $\theta_{\text{PR}}$ around the 111 Bragg angle $\theta_{\text{B}}$ of the diamond phase plate. When we write $\Delta\theta_{\text{PR}}= \theta_{\text{PR}} - \theta_{\text{B}}$, 
the degrees of circular and linear polarizations (Stokes parameters) are expressed as $P_2=\sin (\alpha/\Delta\theta_{\text{PR}})$ and $P_3=-\cos (\alpha/\Delta\theta_{\text{PR}})$, where $\alpha$ is a constant determined experimentally~\cite{SM}.  
The experimental results in Fig. 2B for domain-C and domain-B clearly exhibit opposite asymmetries, indicating that they have opposite magnetic helicities. As shown by the solid lines in the figure, the calculated intensity explains the data well. 
The Fourier components $\bm{m}(\bm{q}_{\text{C}})$ and $\bm{m}(\bm{q}_{\text{B}})$ for domain-C and domain-B, respectively, are related by the [110]-[001] mirror-plane reflection. 
Similar measurements were also performed for other Bragg peaks for domain-A and domain-D. The result of the magnetic helicity measurement is summarized in Fig. 2C by $+$ and $-$ marks. 

The helicities of the four helimagnetic domains perfectly reflect the fourfold and mirror-reflection symmetries of the crystal. 
This result clearly shows that the helicity selection occurs due to the DM-type ASI, depending on the position of the $\bm{q}$ vector. 
When $\bm{q}$ lies exactly on the mirror plane, a cycloidal structure should be selected because the $\bm{D}$ vector is perpendicular to the mirror plane, which is the case for EuIrGe$_3$~\cite{Matsumura22,Kurauchi23,Yambe22}. 
Because of the small value of $\delta_2$ in EuNiGe$_3$, $\bm{q}$ is away from the mirror plane. 
The symmetry constraint of the $\bm{D}$ vector is removed, and consequently, the helical order is realized. 
The ordered moment is also free from the symmetry constraint from the viewpoint of irreducible representation for this $\bm{q}$. 
As a result, the helical plane need not be perpendicular to $\bm{q}$. 
The angle between $\bm{q}$ and the helical plane was experimentally deduced to be $64^{\circ}$~\cite{SM}. 
The distribution of the $\bm{D}$ vector in the reciprocal space, which is expected from the $C_{4v}$ symmetry and the experimentally determined $\bm{m}(\bm{q})$ vectors, is shown schematically in Fig. 2C. 
The angle relation between $\bm{q}$ and $\bm{D}$ changes every 45$^{\circ}$. 
Note that the helical order in EuNiGe$_3$ also accompanies a cycloidal component. 
The helicity of the cycloidal component is fixed by the polar nature of the crystal structure, which is common to all four domains. 
The two types of helicity selection consistent with the crystal symmetry clearly demonstrate that there indeed exists the DM-type ASI in this compound. 

\section{Distorted triangular skyrmion lattice in phase II}
When a magnetic field is applied along the $c$-axis, a phase transition occurs at 2 T. In phase II, the $\bm{q}$ vector exhibits a discontinuous shift, e.g., for domain-C, the original peak at 0 T shown by the open circle in Fig.~\ref{fig3}B jumps to $\bm{q}_{\text{C}1}$ expressed by $(\delta_1, \delta_2)=(0.237,0.072)$. In addition, another Bragg peak simultaneously appears at $\bm{q}_{\text{C}2}$, which is expressed by $(\delta_1^{\;\prime}, \delta_2^{\;\prime})=(0.215, 0.083)$. There also arises another peak at $\bm{q}_{\text{C}3}=-(\bm{q}_{\text{C}1}+\bm{q}_{\text{C}2})$, indicating that the three vectors are related to each other. 
The magnitudes of the $\bm{q}$ vectors are different, and the relative angles are not equal to $120^{\circ}$, as shown in Fig.~\ref{fig3}B. 
Furthermore, the intensities of the three peaks are almost equal. 
Higher-order reflection is also observed at $\bm{q}_1- \bm{q}_3$ with an intensity of $\sim 3$\% of that of the primary peaks. 
These results strongly guarantee that the three peaks form a triple-$\bm{q}$ state and not a multidomain single-$\bm{q}$ state. 

The results of the phase-retarder scan for these three peaks originating from domain-C are shown in Fig. 3A, where the solid lines represent the calculated intensities obtained by using the Fourier components, as shown in each panel. 
These $\bm{m}(\bm{q})$ vectors are determined so that the results of the phase-retarder scans for the three peaks originating from the other domains can be consistently explained as well, where the $\bm{m}(\bm{q}_i)$ ($i=1,2,3$) for different domains are required to be related by the symmetry operations of the crystal. The results of polarization analysis of the diffracted x-rays were also taken into account to deduce the Fourier components~\cite{SM}. 

The most surprising result of these data is that the asymmetry of the data for (4, 0, 0)$-\bm{q}_{\text{C2}}$, which is close to the  $\bm{q}_{\text{B}}$-peak at zero field and is in the \textit{negative} helicity region, is reversed. 
The result at 0 T that the helicity of the helical domain-B is uniquely determined to be \textit{negative} means that the ASI affects the helical order propagating along the direction $(\cos\theta, \sin\theta, 0)$ in the region $\pi < \theta < 5\pi/4$ such that it has a \textit{negative} helicity. 
The $\bm{D}$ vector in this region is antiparallel to the $\bm{q}$ vector. 
Therefore, the Fourier component of the $\bm{q}_{\text{C2}}$ peak in the triple-$\bm{q}$ structure of phase II should be affected by the ASI to have a \textit{negative} helicity. 
However, the observation clearly shows that the helicity of the $\bm{q}_{\text{C2}}$ peak is \textit{positive}. 
This is reflected in the reversal of the sign of $i$ in the $z$ component of $\bm{m}(\bm{q})$. 

With respect to the $\bm{q}_{\text{C3}}$ peak, which is located close to the mirror plane, the ASI is expected to affect the ordering such that it is cycloidal. 
The $\bm{D}$ vector is almost perpendicular to $\bm{q}_{\text{C3}}$. 
However, the $\Delta\theta_{\text{PR}}$-scan data cannot be explained by such a model. The analysis shows that the Fourier component of the $\bm{q}_{\text{C3}}$ peak has a strong helical nature, i.e., the helical plane prefers to be perpendicular to $\bm{q}$, as it is for other peaks of $\bm{q}_{\text{C1}}$ and $\bm{q}_{\text{C2}}$. 
Note that the helicity of the cycloidal component of these peaks, although though much weaker than that of the helical component, does not change with the magnetic-field induced transition from phase I to phase II. It is fixed by the polar nature of the crystal.
Only the helical component changes its helicity when forming the triple-$\bm{q}$ SkL. 

The result of the helicity measurement for the triple-$\bm{q}$ components for domain-D is shown in Fig. 3D, which shows that all the constituent waves have \textit{negative} helicity. This is a perfect mirror reflection of the domain-C structure. 
In this case, the helical component of $\bm{q}_{\text{D2}}$ should be affected by the ASI to have a \textit{positives} helicity since it is close to $\bm{q}_{\text{A}}$ at 0 T and is in the \textit{positive} helicity region. However, the helicity is reversed, and consequently, all three components are unified to have a \textit{negative} helicity to form the triple-$\bm{q}$ SkL. 

The three $\bm{q}$ vectors are away from the symmetric positions in the reciprocal space. They are not related by any symmetry relations. 
Simultaneously, the three Fourier components are not restricted by any symmetry constraints in terms of the irreducible representation for this $\bm{q}$. 
Consequently, the angle between $\bm{q}$ and the helical plane has no symmetry relation; it is $85^{\circ}$ for $\bm{q}_1$, $65^{\circ}$ for $\bm{q}_2$, and $76^{\circ}$ for $\bm{q}_3$. 
The ratio between the $ab$-plane component and the $c$-axis component, $1/\sqrt{m_x^{\;2} + m_y^{\;2}}$, is 0.9 for $\bm{q}_1$, 1.1 for $\bm{q}_2$, and 1.6 for $\bm{q}_3$~\cite{SM}. 

\section{Discussion}
The real-space magnetic structure is described by the superposition of the Fourier components. 
If we neglect the higher-order contributions, then it is expressed by 
\begin{equation}
\bm{M}(\bm{r}) = \sum_{j=1}^{3} [ \bm{m}(\bm{q}_j) \exp \{i(\bm{q}_j\cdot \bm{r} + \phi_j)\} + \text{c.c.} ] \;.
\label{eq:2}
\end{equation}
Since the phase parameters $\phi_j$ cannot be obtained from the diffraction experiment only, we need to assume them to draw a real-space image. 
To describe the skyrmion state, we set them so that the magnetic moment at the center points opposite to the external field. 
The real-space image of the triple-$\bm{q}$ magnetic structure thus obtained is shown in Fig. 3C for domain-C. 
Fig. 3E shows the real-space image obtained for domain-D, which is the mirror reflection image of the domain-C SkL. 
Reflecting the asymmetry of the component $\bm{q}$ vectors, the triangular lattice of the skyrmions is distorted. 
Both structures in Figs. 3C and 3D have the topological skyrmion number of $-1$.  

In phase III above 2.9 T, the $z$ component of $\bm{m}_{\bm{q}}$ vanishes. The magnetic structure is described by a sinusoidal modulation in the $ab$ plane and the uniform magnetization along the $c$-axis. This is a natural result for the spin system that gains more energy from the Zeeman term than from the exchange interactions. 
The $\bm{q}$ vector jumps to $(\delta_1, \delta_2, 0)=(0.25, 0.056, 0)$, the distorted triple-$\bm{q}$ triangular SkL disappears, and the eight Bragg peaks connected by the $C_{4v}$ symmetry operations recover. 
Since this structure does not give rise to an emergent field, the anomalous topological Hall effect should disappear. The result for $\rho_{yx}$ in phase III (Fig. 1D), which seems to be proportional to $\rho_{xx}$, is therefore considered not to be due to the topological Hall effect. 
However, these transport properties need to be studied more carefully. 

The energy gain to form the triangular SkL in EuNiGe$_3$ is larger than the square lattice anisotropy. 
This anisotropy reflecting the fourfold tetragonal symmetry is considered to be caused by the RKKY exchange interaction~\cite{Hayami17,Yambe22}, where the Fermi surface geometry with tetragonal symmetry should play a fundamental role~\cite{Kakihana17}. 
Since the orbital moment of Eu$^{2+}$ is zero, the crystal field anisotropy is expected to be very weak, which is actually reflected in the almost isotropic magnetic susceptibility in the paramagnetic region. 

The energy gain of the SkL formation is also larger than the intrinsic preference for the magnetic helicity due to the ASI. 
In chiral crystals, the helicity requirement to form the SkL, i.e., the requirement that the helicities of the three $\bm{q}$ components need to be the same, is automatically fulfilled~\cite{Muhlbauer09,Yu10}. The SkL formation is assisted by the intrinsic ASI existing in the system. 
In centrosymmetric compounds, where both helicities are equivalently allowed, the spin system spontaneously selects one helicity when forming the SkL. 
In the present case of EuNiGe$_3$, where the intrinsic helicity of $\bm{m}(\bm{q})$ changes every $45^{\circ}$ in reciprocal space, the helicity of one of the triple-$\bm{q}$ components is unavoidably reversed. 

The present experiment confirms that all three constituent waves of the magnetic spiral of the triangular SkL are unified to have the same helicity to minimize the total free energy of the spin system. This occurs even when one of the triple-$\bm{q}$ components has the opposite intrinsic helicity. 
Although the helicity of the primary $\bm{q}_1$ component is determined by the intrinsic ASI of the polar tetragonal structure, the helicities of the other two components are unified to that of the primary component. 
Therefore, the driving force to form such an emergent state is not necessarily the DM-type ASI intrinsic to the noncentrosymmetric crystal structure. 
It is more likely to be associated with the competing or higher-order exchange interactions, as has been theoretically investigated~\cite{Hayami17,Okubo12,Leonov15}. 
Moreover, the total free energy of the spin system is minimized when the SkL becomes triangular, where the skyrmions are closely packed in two dimensions, even in the square lattice environment with strong coupling with fourfold symmetry. 
Since the two symmetries are not compatible, the SkL spontaneously deforms into an asymmetric structure.  

\vspace{5mm}
\subsection*{acknowledgments}
The authors acknowledge valuable discussions with A. Tanaka and A. O. Leonov. 
This work was supported by JSPS Grant-in-Aid for Scientific Research (B) (No. JP20H01854) and by JSPS Grant-in-Aid for Transformative Research Areas (Asymmetric Quantum Matters, 23A202, No. JP23H04867). 
The synchrotron experiments were performed under the approval of the Photon Factory Program Advisory Committee (Nos. 2020G034 and 2022G114). 
MT is supported by JST, the establishment of university fellowships toward the creation of science technology innovation, Grant No. JPMJFS2129.

\bibliography{EuNiGe3S-1}

\begin{figure*}
\begin{center}
\includegraphics[width=12cm]{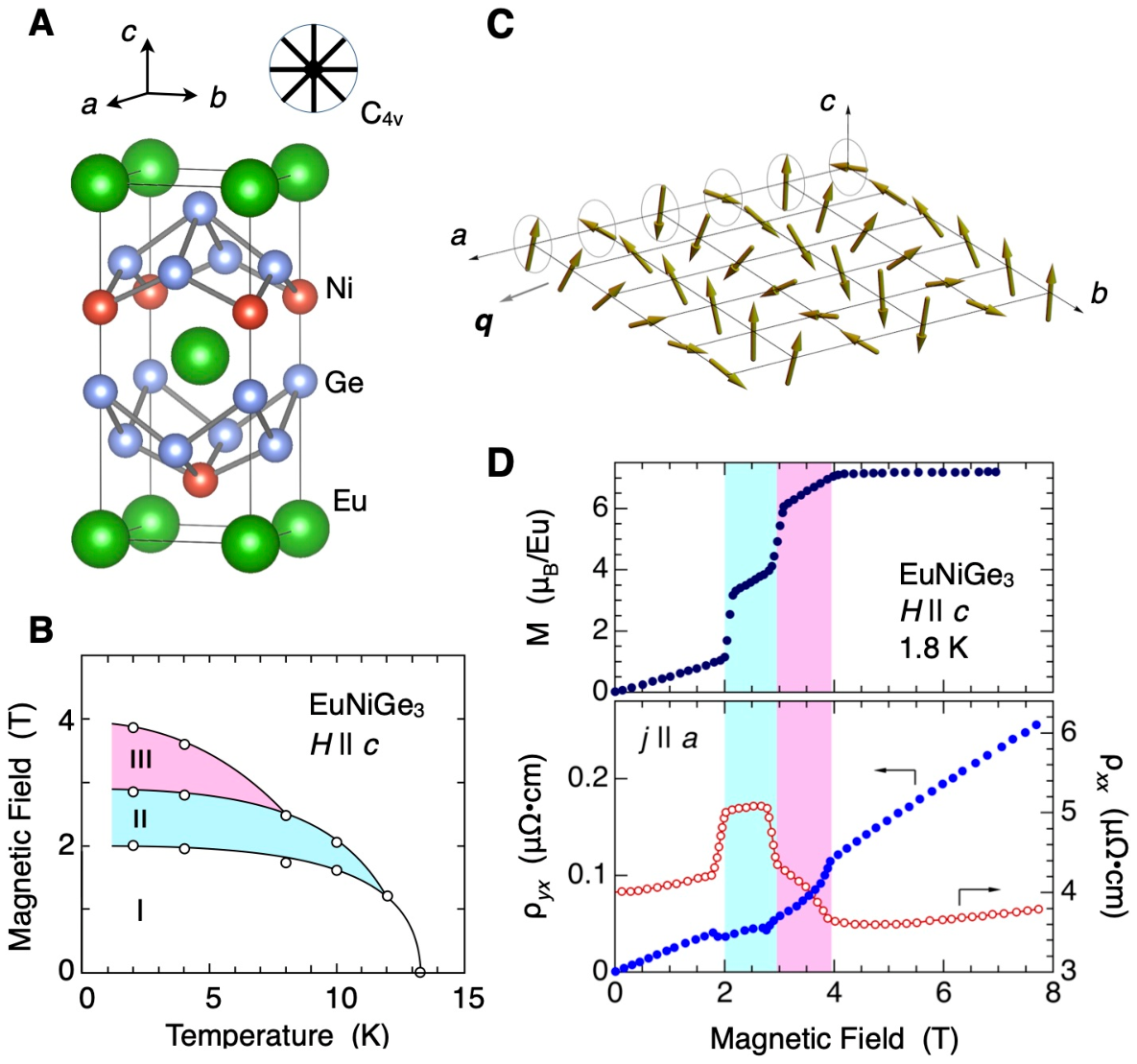}
\caption{
Structures and bulk properties of EuNiGe$_3$. 
(A) Body centered tetragonal lattice of EuNiGe$_3$ with a fourfold axis and the mirror planes including the $c$-axis (point group $C_{4v}$). 
(B) Magnetic phase diagram for $H \parallel c$ constructed from the bulk property measurements~\cite{Maurya14,Kakihana17}. 
(C) Helical magnetic structure at zero magnetic field described by a single propagation vector. The amplitude along the $c$-axis is larger than those in the $ab$ plane. The helical plane is not perpendicular to $\bm{q}$. 
(D) Magnetization ($M$), magnetoresistance ($\rho_{xx}$), and Hall resistivity ($\rho_{yx}$) for $H \parallel c$ at 1.8 K for the field-increasing process~\cite{Iha20}. 
Phase transitions occur at 2 T, 2.9 T, and 3.9 T, above which a fully polarized ferromagnetic state is realized. 
The low-field phase below 2 T, the intermediate phase between 2 T and 2.9 T, and the high-field phase between 2.9 T and 3.9 T are labeled phases I,  II, and  III, respectively. 
}
\label{fig1}
\end{center}
\end{figure*}

\begin{figure*}
\begin{center}
\includegraphics[width=12cm]{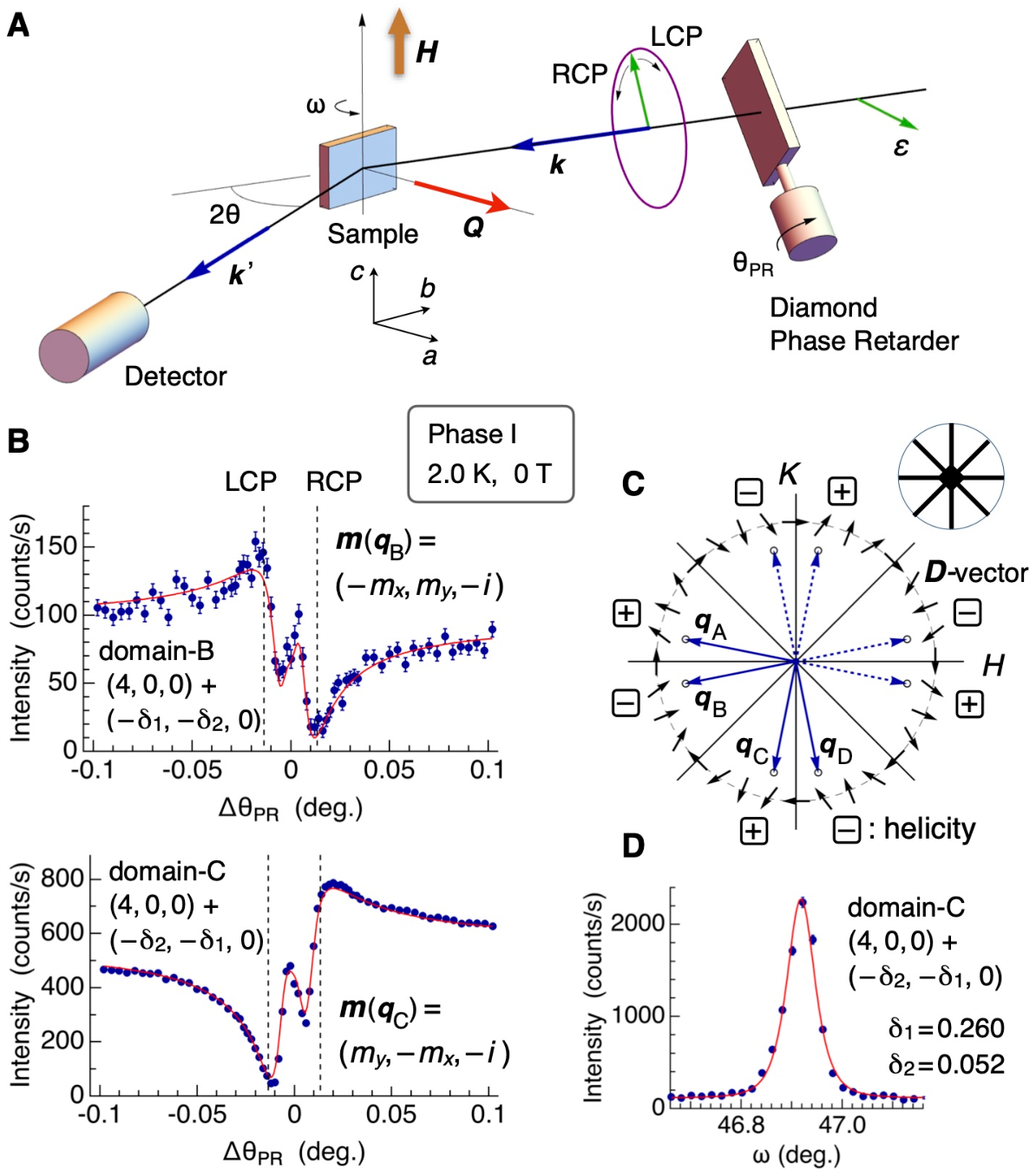}
\caption{
Single helicity of the helimagnetic structure at zero field confirmed by resonant x-ray diffraction (RXD). 
(A) Experimental setup of RXD in our study. The linear $\pi$-polarization of the incident beam from the synchrotron source is tuned to circular polarization by using a diamond phase retarder. A vertical magnetic field is applied on the sample. 
(B) The phase-retarder angle ($\Delta \theta_{\text{PR}}$) dependences of the magnetic Bragg peak intensity at $(4, 0, 0)+(-\delta_1, -\delta_2, 0)$ and $(4, 0, 0)+(-\delta_2, -\delta_1, 0)$, corresponding to the single-$\bm{q}$ helimagnetic domain-B and -C, respectively, where $\delta_1=0.260$ and $\delta_2=0.052$. 
The vertical dashed lines indicate the positions of left- or right-handed circular polarization. At large $\Delta \theta_{\text{PR}}$ values the polarization becomes elliptic and approaches linear $\pi$-polarization. The solid lines are the calculated intensities for the Fourier component $\bm{m}(\bm{q})$ with $(m_x, m_y)=(0.395, 0.526)$.  
(C) The experimentally determined magnetic helicities of the four domains are indicated by $+$ and $-$ signs, which perfectly reflect the four fold and mirror reflection symmetries of the crystal. The short arrows schematically represent the $\bm{q}$-dependent $\bm{D}$ vectors estimated from the $\bm{m}(\bm{q})$ measurements and symmetry considerations. 
(D) Example of the magnetic Bragg peak measured at a resonance energy of 7.612 keV. 
}
\label{fig2}
\end{center}
\end{figure*}

\begin{figure*}
\begin{center}
\includegraphics[width=18cm]{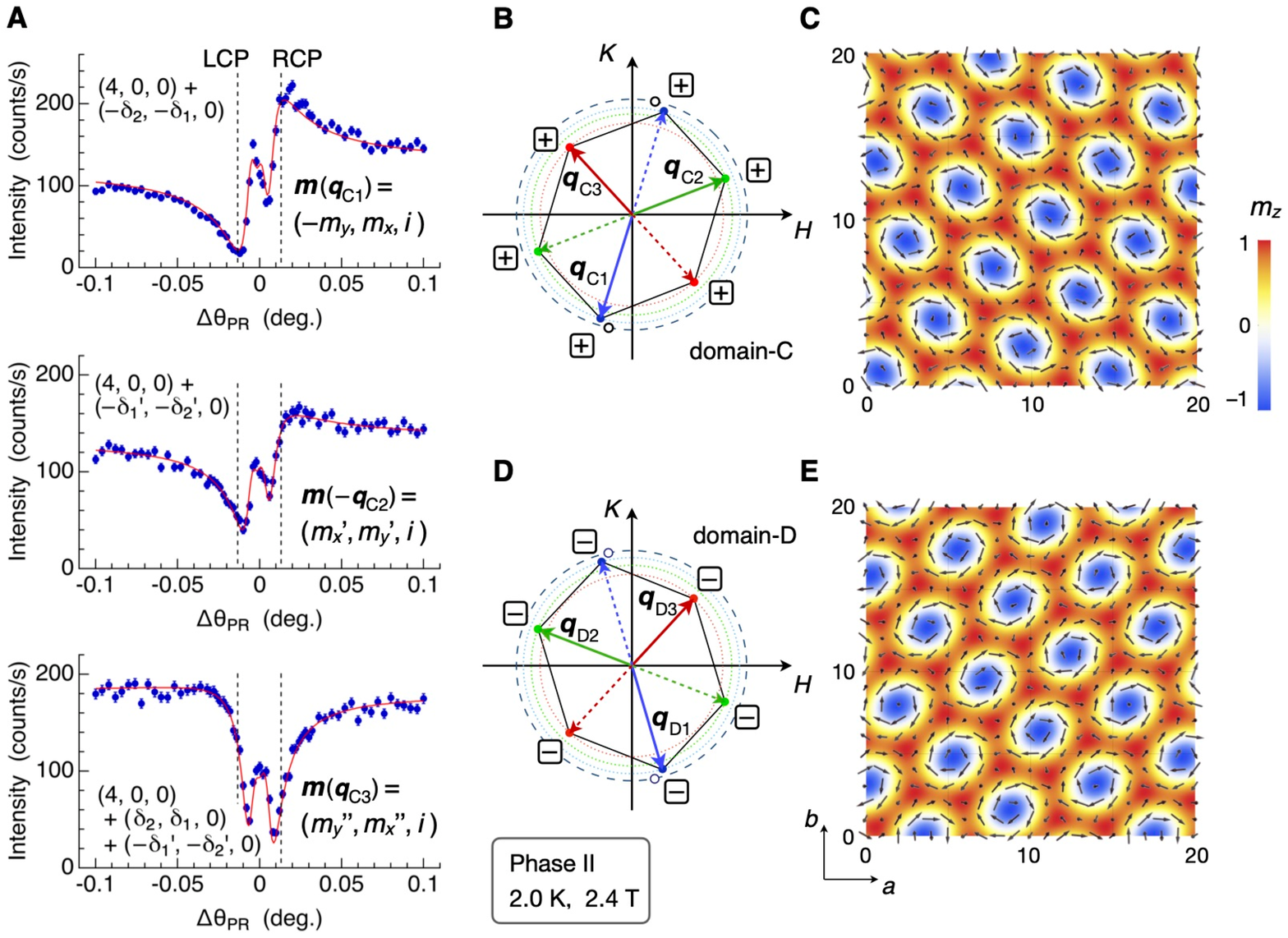}
\caption{
Observation of the triangular skyrmion lattice state described by triple-$\bm{q}$ Fourier components. 
(A) Results of the $\Delta \theta_{\text{PR}}$ dependence of the Bragg peak intensity at three constituent $\bm{q}$ vectors for domain-C, where $(\delta_1, \delta_2)=(0.237, 0.072)$ for the primary peak at $\bm{q}_1$,  
$(\delta_1^{\prime}, \delta_2^{\prime})=(0.215, 0.083)$ for the secondary peak at $\bm{q}_2$, and the third peak appears at $\bm{q}_3=-(\bm{q}_1+\bm{q}_2)$. 
The solid lines are the calculated intensities for the Fourier components with $(m_x, m_y)=(0.41, 1.03)$ for $\bm{q}_1$, $(m_x^{\prime}, m_y^{\prime})=(0.90, 0.09)$ for $\bm{q}_2$, and $(m_x^{\prime\prime}, m_y^{\prime\prime})=(0.28, 0.56)$ for $\bm{q}_3$. 
All three components have the same helicity (+) as shown in (B).  
The open circle near $\bm{q}_1$ is the primary $\bm{q}$ vector at zero field. 
(C) Real-space image of the magnetic structure in phase II for domain-C in the $ab$ plane. The color and the arrows represent the $z$ and the $xy$ components, respectively.  
(D,E) Results for the domain-D peaks, where all the data are explained by the same helicity ($-$).  The-real space image is the mirror reflection of that for domain-C. 
}
\label{fig3}
\end{center}
\end{figure*}


\clearpage
\begin{widetext}
\begin{center}
\textbf{\large{Supplemental Material}}
\end{center}
\vspace{5mm}

\begin{center}
\textbf{\large{Distorted triangular skyrmion lattice in a noncentrosymmetric tetragonal magnet}} \\
\vspace{2mm}
T. Matsumura, K. Kurauchi, M. Tsukagoshi, N. Higa, H. Nakao, M. Kakihana, M. Hedo, T. Nakama, and Y. \={O}nuki
\end{center}
\vspace{20mm}
\end{widetext}

\renewcommand{\topfraction}{1.0}
\renewcommand{\bottomfraction}{1.0}
\renewcommand{\dbltopfraction}{1.0}
\renewcommand{\textfraction}{0.01}
\renewcommand{\floatpagefraction}{1.0}
\renewcommand{\dblfloatpagefraction}{1.0}
\setcounter{topnumber}{5}
\setcounter{bottomnumber}{5}
\setcounter{totalnumber}{10}

\renewcommand{\theequation}{S\arabic{equation}}
\renewcommand{\thefigure}{S\arabic{figure}}
\renewcommand{\thetable}{S-\Roman{table}}
\setcounter{section}{0}
\setcounter{equation}{0}
\setcounter{figure}{0}
\setcounter{page}{1}

\section{Experiment}
\subsection{Sample}
The EuNiGe$_3$ single crystal used in the resonant X-ray diffraction experiment was grown by the In-flux method as described in Ref.~\onlinecite{Kakihana17}. 
The starting elemental materials were placed in an alumina crucible, which was encapsulated in a quartz ampoule. 
The ampoule was heated up to 1130~$^{\circ}$C, held for three days, and cooled down to 500~$^{\circ}$C by taking 15 days in total. 
The In flux was removed at 250~$^{\circ}$C by spinning the ampoule in a centrifuge. 
The lattice parameter, electrical resistivity, specific heat, magnetic susceptibility, and magnetization are reported in Ref.~\onlinecite{Kakihana17}, which are all consistent with the previous data by Maurya et al. reported in Ref. \onlinecite{Maurya14}.

\subsection{Resonant X-ray Diffraction}
\begin{figure}
\begin{center}
\includegraphics[width=8cm]{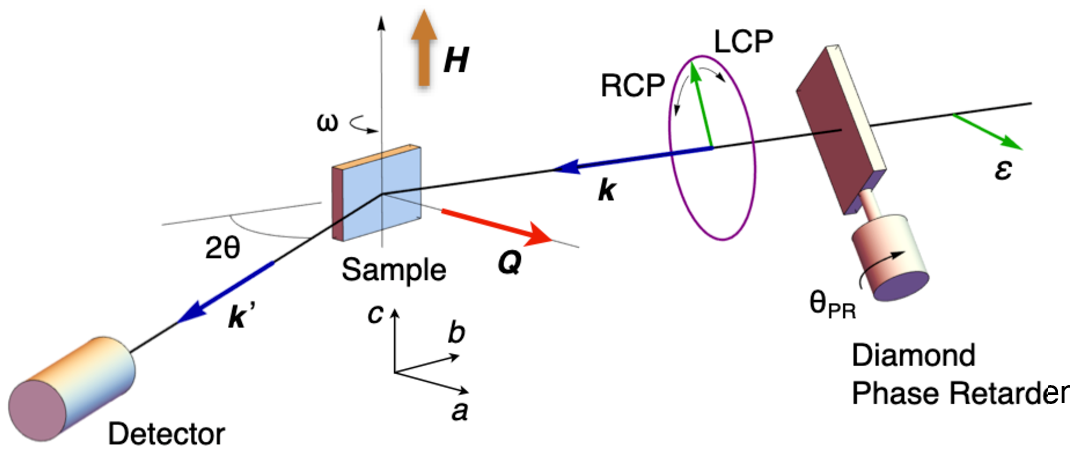}
\caption{Scattering geometry of the resonant X-ray diffraction experiment with a phase retarder system inserted in the incident beam. 
}
\label{figS:config}
\end{center}
\end{figure}

Resonant x-ray diffraction (RXD) experiment was performed at BL-3A of the Photon Factory, KEK, Japan. 
Figure~\ref{figS:config} shows the scattering geometry of the RXD experiment. 
A plate-shaped sample with a mirror polished (100)-plane surface, $1.9 \times 1.2$ mm$^2$ in area and 0.2 mm in thickness, was mounted in a vertical field 8 T superconducting cryomagnet so that the [001]-axis ($c$-axis) was vertical and the (100)-plane was normal to the scattering vector $\bm{Q}=\bm{k}' - \bm{k}$ for the $(H 0 0)$ reflection. 
We used X-ray energies around the $L_2$ absorption edge of Eu. 

Circularly polarized beam was obtained by using a diamond phase-retarder system. 
When the incident X-ray from the synchrotron source, which is polarized in the horizontal plane ($\pi$-polarization), passes through the diamond phase plate set near a Bragg angle, a phase difference occurs between the $\sigma$ and $\pi$ components with respect to the scattering plane tilted by $45^{\circ}$~\cite{Hirano95}.
The phase difference is proportional to $1/(\theta_{\text{PR}} - \theta_{\text{B}})$, where $\theta_{\text{B}}$ is the Bragg angle of the phase plate. 
It is therefore possible to tune the incident linear polarization to right-handed circular polarization (RCP) and left-handed circular polarization (LCP) by manipulating $\Delta\theta_{\text{PR}} = \theta_{\text{PR}} - \theta_{\text{B}}$ around the Bragg angle $\theta_{\text{B}}$. 
Here in this experiment, we used a 111 Bragg reflection of a diamond phase plate with a thickness of 0.63 mm. 
The polarization state of the X-ray after transmitting the phase plate can be described by using the Stokes parameters $P_2$ and $P_3$, where $P_2$ represents the degree of circular polarization ($+1$ for RCP and $-1$ for LCP) and $P_3$ represents the degree of linear polarization ($+1$ for $\sigma$ and $-1$ for $\pi$ polarization)~\cite{Lovesey96}. 
In the horizontal scattering-plane geometry in our experiment, $P_2$ and $P_3$ can be expressed as $P_2= \sin (\alpha/\Delta\theta_{\text{PR}})$ and $P_3 = -\cos (\alpha/\Delta\theta_{\text{PR}})$, where $\alpha$ is an experimentally determined parameter specific to the phase plate. 
Near $\Delta\theta_{\text{PR}} = 0$ the beam becomes depolarized. 
$P_1$ ($+1$ for $45^{\circ}$ and $-1$ for $-45^{\circ}$ linear polarization) is zero in the present setup.

We use the scattering-amplitude-operator method to analyze the experimental results~\cite{Lovesey96}. 
The resonant scattering amplitude can be expressed by a $2\times 2$ matrix $\hat{F}$, consisting of four elements of the scattering amplitude for 
$\sigma$-$\sigma'$, $\pi$-$\sigma'$, $\sigma$-$\pi'$, and $\pi$-$\pi'$:
\begin{equation}
\hat{F} = \begin{pmatrix} F_{\sigma\sigma'} & F_{\pi\sigma'} \\ 
F_{\sigma\pi'} & F_{\pi\pi'} \end{pmatrix}\,.
\label{eq:scampG}
\end{equation}
Using the four elements of (\ref{eq:scampG}), the scattering intensity can be written as 
\begin{align}
I &= 
\frac{1}{2} \bigl(\, |F_{\sigma\sigma'}|^2 + |F_{\sigma\pi'}|^2 + |F_{\pi\sigma'}|^2 + |F_{\pi\pi'}|^2 \,\bigr) \nonumber \\
 &\;\;\;\; + P_1 \text{Re} \bigl\{\, F_{\pi\sigma'}^*F_{\sigma\sigma'} + F_{\pi\pi'}^*F_{\sigma\pi'} \,\bigr\} \nonumber \\
 &\;\;\;\; + P_2 \text{Im} \bigl\{\, F_{\pi\sigma'}^*F_{\sigma\sigma'} + F_{\pi\pi'}^*F_{\sigma\pi'} \,\bigr\} 
 \label{eq:CrossSec1} \\
 &\;\;\;\; +  \frac{1}{2} P_3\bigl(\, |F_{\sigma\sigma'}|^2 + |F_{\sigma\pi'}|^2 - |F_{\pi\sigma'}|^2 - |F_{\pi\pi'}|^2 \,\bigr) 
 \,. \nonumber
\end{align}
Therefore, the intensity for the incident beam described by the Stokes parameters $(P_1, P_2, P_3)$ can generally be written as
\begin{equation}
I = C_0 + C_1 P_1 + C_2 P_2 + C_3 P_3 \,,
\label{eq:CrossSec2}
\end{equation}
which can be used as a fitting function for the $\Delta\theta_{\text{PR}}$ scan with four parameters of $C_n$ ($n=0\sim 3$)~\cite{Matsumura17}. 

For the $E1$ resonance, the scattering amplitude from a magnetic dipole order is expressed as   
\begin{equation}
F_{\varepsilon\varepsilon'}(\omega) =  f(\omega) ( \bm{\varepsilon}' \times \bm{\varepsilon}) \cdot   \bm{Z}_{\text{M}} \;, 
\label{eq:Freso}
\end{equation}
where
\begin{equation}
\bm{Z}_{\text{M}} = \sum_j \bm{M}(\bm{r}_j) \exp (-i \bm{Q}\cdot\bm{r}_j) \;, 
\end{equation}
represents the magnetic dipole structure factor for the scattering vector $\bm{Q} = \bm{k}' - \bm{k}$, $\bm{m}_j$ the magnetic dipole moment located at $\bm{r}_j$, and $f(\omega)$ the spectral function for the $E1$ resonance~\cite{Nagao05,Nagao06}. 
When the magnetic structure is expressed by  
\begin{equation}
\bm{M}(\bm{r}_j) = \bm{m}(\bm{q}) \exp (i\bm{q}\cdot \bm{r}_j) + \bm{m}(\bm{q})^{*} \exp (-i\bm{q} \cdot \bm{r}_j) \;, 
\label{eq:magr}
\end{equation}
$\bm{Z}_{\text{M}}$ is equal to $\bm{m}(\bm{q})$ and $\bm{m}(\bm{q})^{*}$ when $\bm{Q}=\bm{G}+\bm{q}$ and $\bm{Q}=\bm{G}-\bm{q}$, respectively, where $\bm{G}$ represents a reciprocal lattice vector. 

\begin{figure}
\begin{center}
\includegraphics[width=8.5cm]{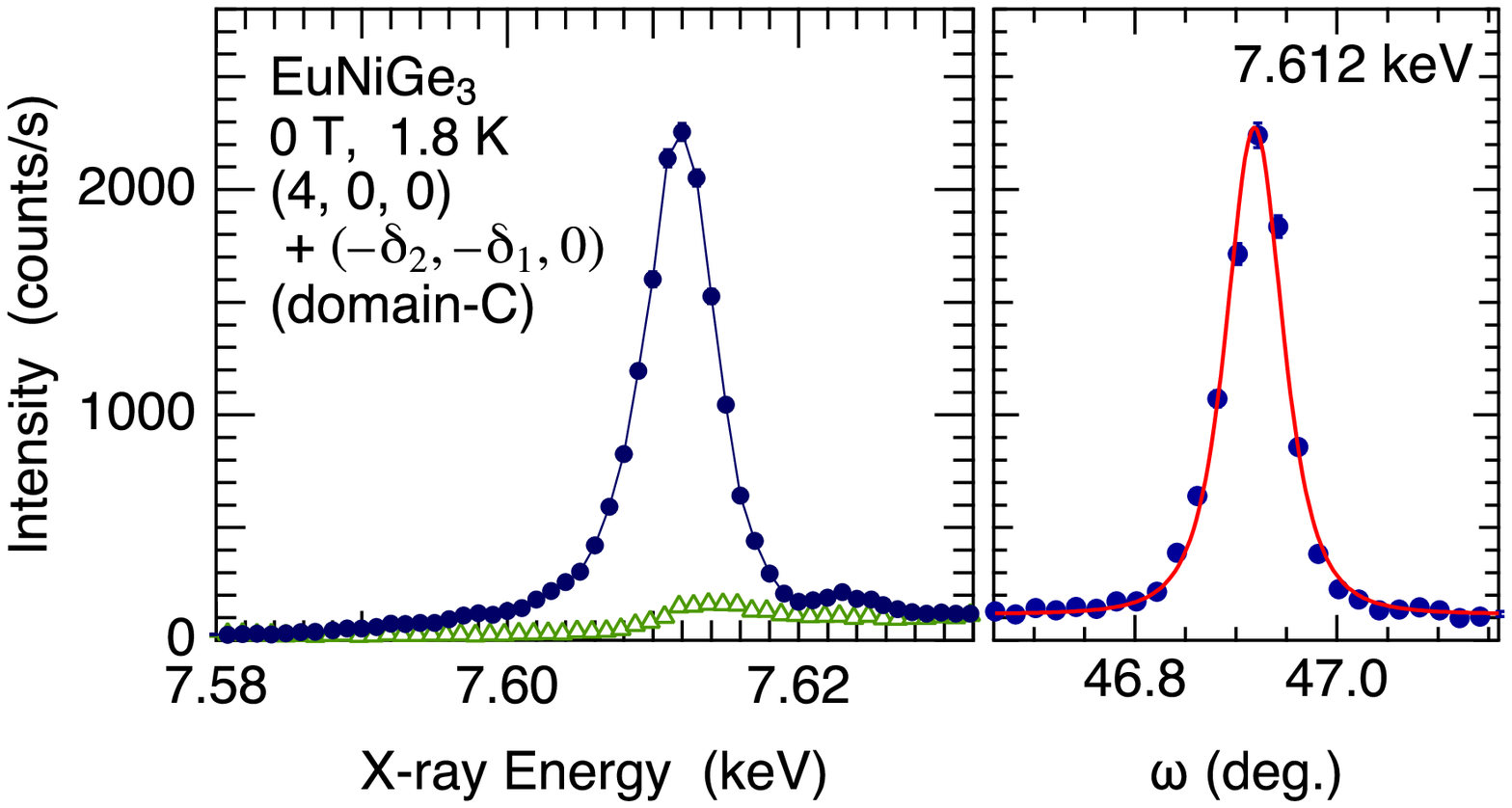}
\caption{(left) X-ray energy dependence of the $(4, 0, 0)+(-\delta_2, -\delta_1, 0)$ Bragg peak at 0 T and 1.8 K without phase retarder. 
$(\delta_1, \delta_2)=(0.26, 0.052)$. 
(right) Rocking curve of the Bragg peak at the resonance energy of 7.612 keV. }
\label{figS:Escan}
\end{center}
\end{figure}

Figure \ref{figS:Escan} shows the X-ray energy dependence of the magnetic Bragg reflection at $(4, 0, 0)+(-\delta_2, -\delta_1, 0)$, corresponding to the domain-C. The intensity exhibits a resonant enhancement centered at 7.612 keV.  
Since this energy corresponds to the $L_2$ absorption edge of Eu, the signal directly reflects the ordered magnetic moment of the Eu ion.

\begin{figure}
\begin{center}
\includegraphics[width=8cm]{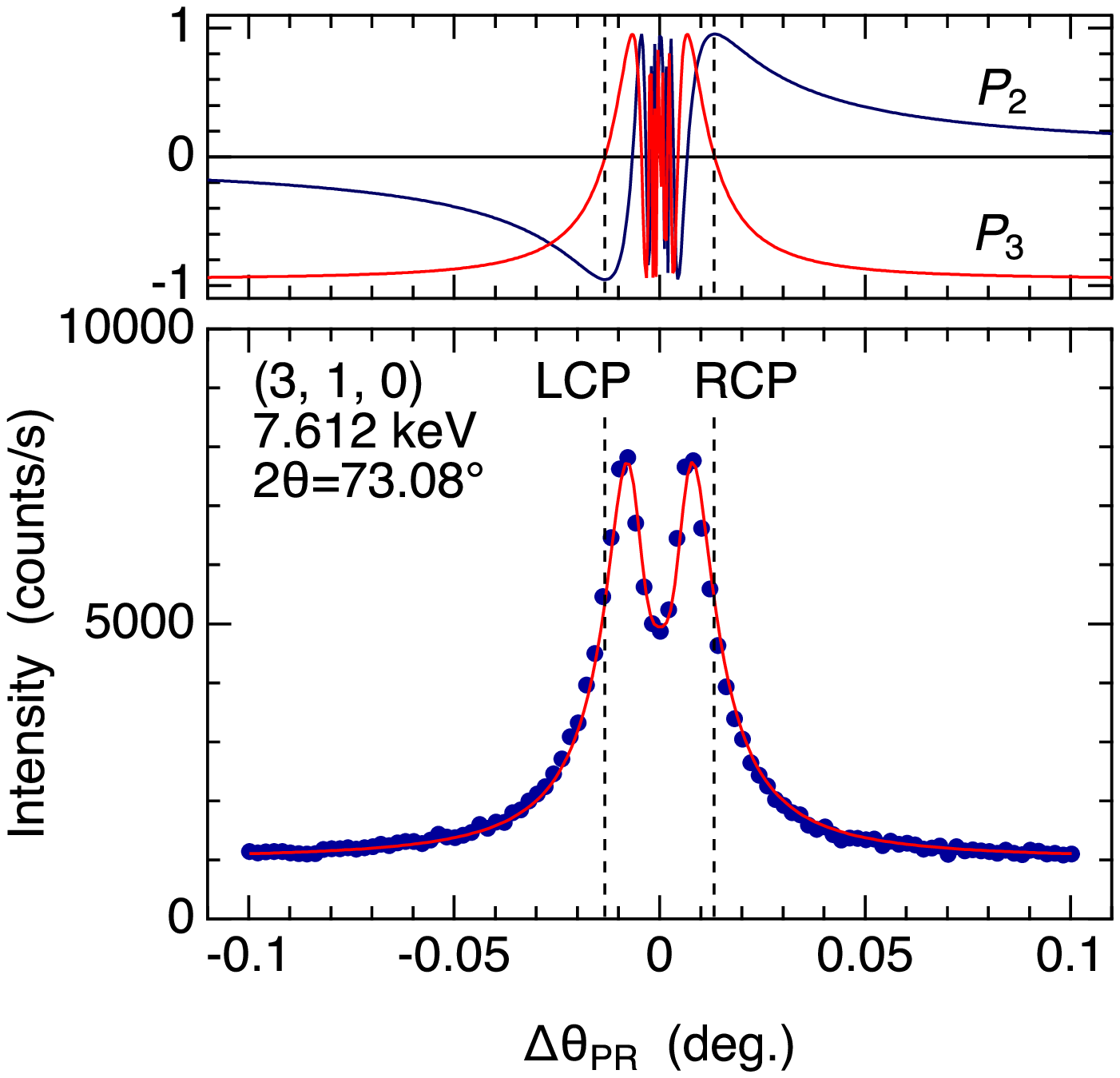}
\caption{(top) $\Delta\theta_{\text{PR}}$ dependence of the Stokes parameters $P_2$ and $P_3$, which are expressed as $P_2= \sin (\alpha/\Delta\theta_{\text{PR}})$ and $P_3 = -\cos (\alpha/\Delta\theta_{\text{PR}})$. 
(bottom) $\Delta\theta_{\text{PR}}$ dependence of the intensity of the $(3, 1, 0)$ fundamental Bragg reflection. 
Solid line is a fit with a convolution of a Gaussian resolution function, from which $\alpha=0.02094$ was obtained. 
The vertical dashed lines represent the positions of the offset angle $\Delta\theta_{\text{PR}}$ where the RCP and LCP states are obtained. 
}
\label{figS:PRth310}
\end{center}
\end{figure}

Figure \ref{figS:PRth310} shows the $\Delta\theta_{\text{PR}}$ dependence of the intensity of the $(3, 1, 0)$ fundamental Bragg reflection.
Since this intensity is purely due to the Thomson scattering, which can be expressed by $F_{\sigma\sigma'}=1$ and $F_{\pi\pi'}=\cos 2\theta$, 
the intensity is expressed as
\begin{equation}
I \propto \Bigl( 1-\frac{1-P_3}{2} \sin^2 2\theta \Bigr) \;.
\end{equation}
By fitting the data with this function, the phase-plate parameter $\alpha= 0.02094$ (deg.) was deduced. 
We use this $\alpha$ for further analyses of the magnetic signals. Convolution with a Gaussian resolution function is taken into account in the analyses. 
When $|\Delta\theta_{\text{PR}}| > 0.1^{\circ}$, $P_3$ exceeds over 0.95 as shown in the top panel. In this region, the beam is almost linearly $\pi$-polarized. 
Around $\Delta\theta_{\text{PR}}\sim 0$, the beam is depolarized. 

\newpage
\section{Data Analysis}
\subsection{magnetic structure at zero field}
\begin{figure}
\begin{center}
\includegraphics[width=8.5cm]{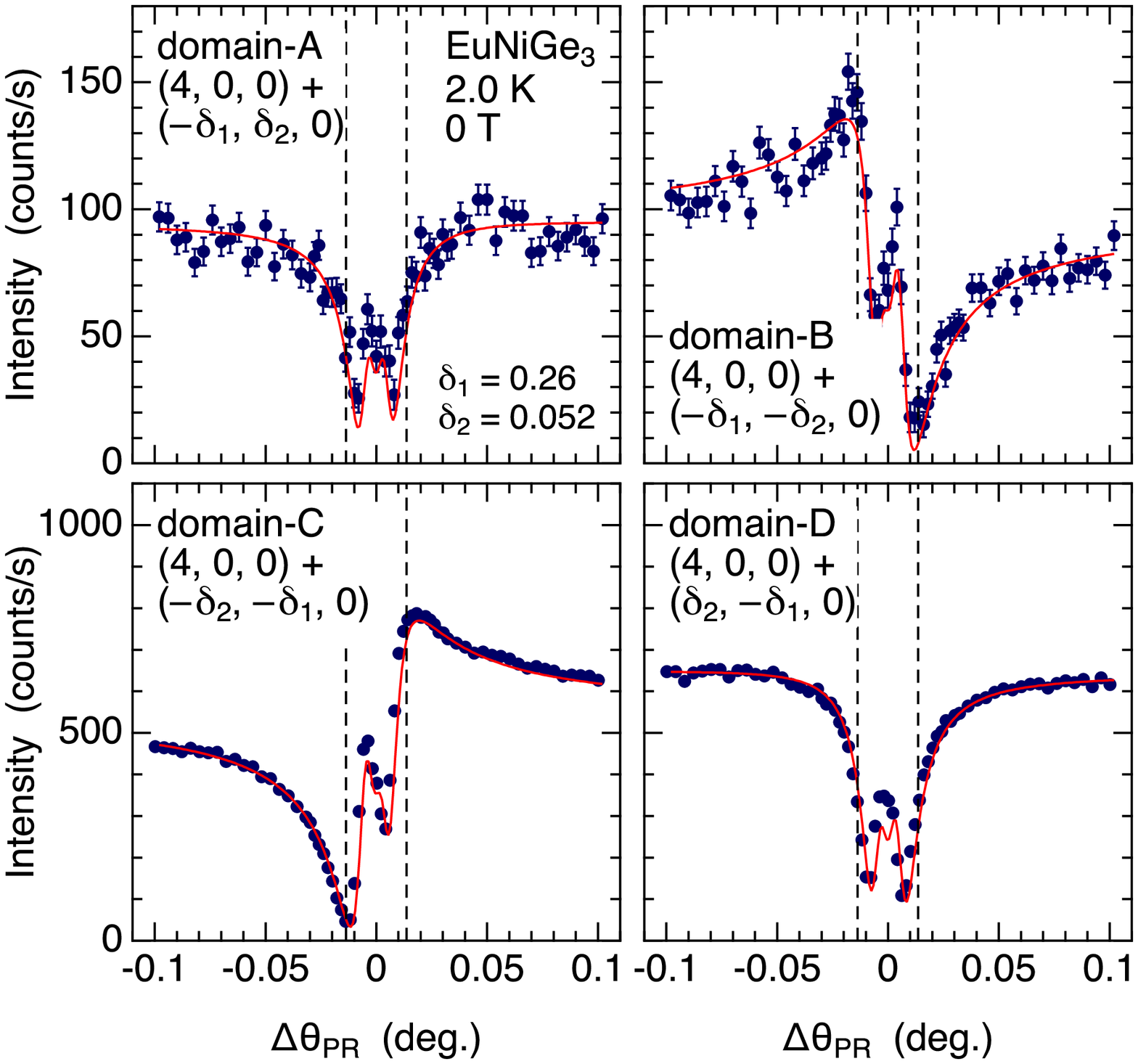}
\caption{$\Delta\theta_{\text{PR}}$ dependences of the magnetic Bragg-peak intensities around the (4, 0, 0) fundamental reflection. 
The x-ray energy is 7.612 keV at resonance. The background has been subtracted. 
The solid lines are the calculated intensity curves for the Fourier components summarized in Table~\ref{tbl:mq24kG}. 
}
\label{figS:PRthscans0T}
\end{center}
\end{figure}

Figure \ref{figS:PRthscans0T} shows the $\Delta\theta_{\text{PR}}$ dependences of the intensity of the resonant magnetic Bragg reflections around the (4, 0, 0) fundamental reflection from the lattice. The four data sets correspond to the four domains, A, B, C, and D, shown in Fig. 2 of the main text. 
The solid lines are the calculated intensities obtained from Eq.~(\ref{eq:CrossSec1}) by assuming the helical magnetic structures for the respective domains. 

Clear asymmetric intensity variation is observed for the domains B and C, whereas the asymmetry is weak for the domains A and D. 
This is because of the geometrical reason associated with the factor $(\bm{\varepsilon}^{\prime} \times \bm{\varepsilon}) \cdot \bm{Z}_{\text{m}}$ in Eq. (\ref{eq:Freso}). 
The asymmetry is due to the $C_2 P_2$ term in Eq. (\ref{eq:CrossSec2}), which arises from $F_{\pi\pi'}^{\;*}F_{\sigma\pi'}$ as expressed in Eq.~(\ref{eq:CrossSec1}). Note that $F_{\sigma\sigma'}=0$ for the $E1$ resonance from magnetic dipole moment since $(\bm{\varepsilon}^{\prime} \times \bm{\varepsilon})=0$.  
Here, $F_{\pi\pi'}$ is almost equal for all the four domains because $F_{\pi\pi'}$ is proportional to the $c$-axis component of $\bm{m}(\bm{q})$. 
On the other hand, $F_{\sigma\pi'}$, which is associated with the $ab$-plane component of $\bm{m}(\bm{q})$, is large for the B and C domains, whereas it is small for the A and D domains due to the geometrical relation between $\bm{m}(\bm{q})$ and $\bm{\varepsilon}^{\prime} \times \bm{\varepsilon}$. 

\begin{figure}
\begin{center}
\includegraphics[width=8.5cm]{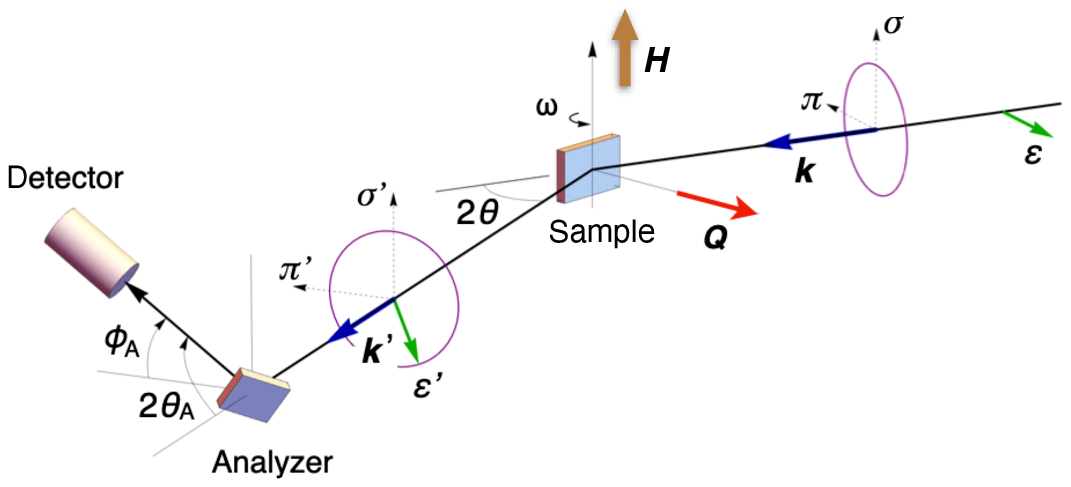}
\caption{Scattering geometry of linear polarization analysis of the diffracted beam. Phase retarder is removed from the incident beam path. 
The incident beam is $\pi$-polarized. 
}
\label{figS:ScattConfigANA}
\end{center}
\end{figure}

The ratio between the $ab$-plane and the $c$-axis components in $\bm{m}(\bm{q})$ is more sensitively reflected in the linear polarization analysis of the diffracted beam. 
The scattering geometry of this analysis is shown in Fig.~\ref{figS:ScattConfigANA}. 
We used a 006 Bragg reflection of a pyrolytic graphite (PG) analyzer crystal, where the angle $2\theta_{\text{A}}$ is $93.5^{\circ}$ at the resonance energy of 7.612 keV in this experiment. 
This $2\theta_{\text{A}}$ angle is close to $90^{\circ}$ and effectively eliminate the $\pi'$ ($\sigma'$) polarization at $\phi_{\text{A}}=0^{\circ}$ ($90^{\circ}$). 

\begin{figure}[t]
\begin{center}
\includegraphics[width=8.5cm]{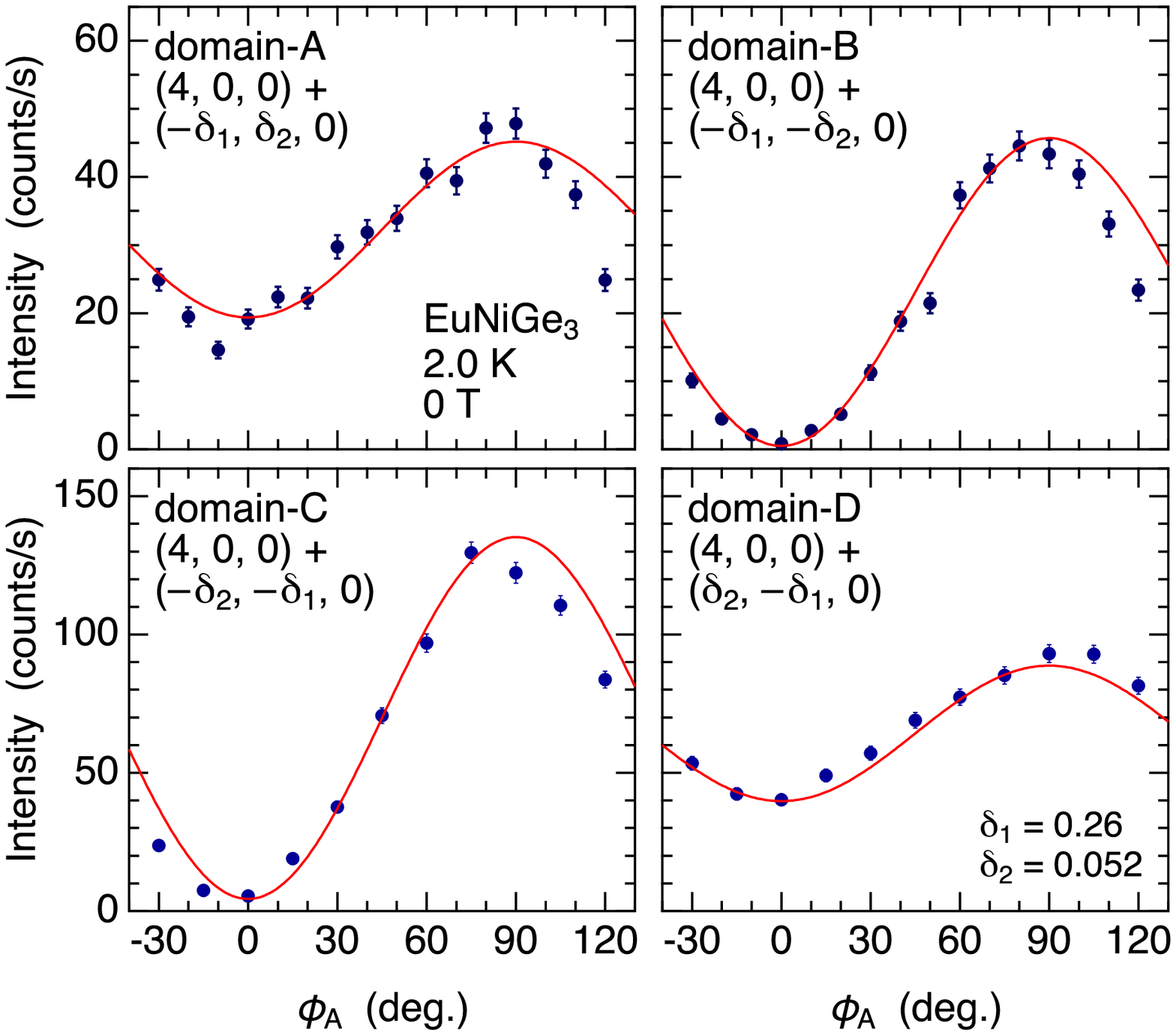}
\caption{Analyzer angle ($\phi_{\text{A}}$) dependences of the magnetic Bragg-peak intensities around the (4, 0, 0) fundamental reflection at 0 T. 
The x-ray energy is 7.612 keV at resonance. 
The solid lines are the calculated intensity curves for the Fourier components summarized in Table~\ref{tbl:mq0T}. 
}
\label{figS:POLscans0T}
\end{center}
\end{figure}

The results of the linear polarization analysis for the four magnetic domains are shown in Fig.~\ref{figS:POLscans0T}. 
In this polarization analysis, the intensity is proportional to $|F_{\pi\sigma'}|^2$ ($|F_{\pi\pi'}|^2$) when $\phi_{\text{A}}=0^{\circ}$ ($\phi_{\text{A}}=90^{\circ}$). Therefore, this analysis is complementary to the $\Delta\theta_{\text{PR}}$ dependence measurement, which is sensitive to $F_{\pi\pi'}^{\;*}F_{\sigma\pi'}$. 
All the four data sets in Fig.~\ref{figS:POLscans0T} show finite intensity for $\pi$-$\pi'$ ($\phi_{\text{A}}=90^{\circ}$), indicating that the $c$-axis component commonly exists in $\bm{m}(\bm{q})$. 
The $\pi$-$\sigma'$ intensity at $\phi_{\text{A}}=0^{\circ}$, on the other hand, is finite for the domains A and D and is very weak for the domains B and C. This is also due to the geometrical relation between $\bm{m}(\bm{q})$ and $\bm{\varepsilon}^{\prime} \times \bm{\varepsilon}$. 
By combining all the data sets in Figs.~\ref{figS:PRthscans0T} and \ref{figS:POLscans0T}, we determined the Fourier component $\bm{m}(\bm{q})$ for the four magnetic domains as summarized in Table~\ref{tbl:mq0T}. 
This is a helimagnetic structure with an elliptic helical plane elongated along the $c$-axis as concluded by neutron diffraction~\cite{Fabreges16}. 
In addition, more detailed analysis in this work show that the helical plane is not perpendicular to the $\bm{q}$ vector and makes an angle $64^{\circ}\pm 2^{\circ}$. 
The real space structure is shown in Fig. 1 of the main text. 
The calculated intensities of the $\Delta\theta_{\text{PR}}$ scan and the $\phi_{\text{A}}$ scan for the four domains of this magnetic structure are shown by the solid lines in Figs.~\ref{figS:PRthscans0T} and \ref{figS:POLscans0T}, which reproduce the experimental data well. 

\begin{table}
\caption{Fourier components of the helical magnetic structure at zero field for the four domains. 
The parameters are $(\delta_1, \delta_2)=(0.26, 0.052)$ and $(m_x, m_y)=(0.40, 0.53)$. 
The angle between $\bm{q}$ and $\bm{m}(\bm{q})$ is $64^{\circ} \pm 2^{\circ}$.  
The ratio between the $z$ ($c$-axis) and the $xy$ ($ab$-plane) components of $\bm{m}(\bm{q})$, $1/\sqrt{m_x^{\;2} + m_y^{\;2}}$, is $1.9 \pm 0.1$. }
\label{tbl:mq0T}
\begin{tabular}{cccc}
\hline
domain & $\bm{q}$ & $\bm{m}(\bm{q})$ & helicity \\
\hline
A & $(-\delta_1, \delta_2, 0)$ & $( m_x, m_y, i)$  & $+$ \\
B & $(-\delta_1, -\delta_2, 0)$ & $( m_x, -m_y, i)$ & $-$  \\
C & $(-\delta_2, -\delta_1, 0)$ & $(-m_y, m_x, i)$ & $+$  \\
D & $( \delta_2, -\delta_1, 0)$ & $( m_y,  m_x,  i)$ & $-$ \\
\hline
\end{tabular}
\end{table}

As explained in the main text, the helimagnetic structures for the four domains perfectly reflect the $C_{4v}$ crystal symmetry. 
The magnetic structure of the domain A is transformed to that of domain C by the $90^{\circ}$ rotation and to that of domain D by the mirror  reflection with respect to the [110]--[001] plane. 
The magnetic structure of the domain C is transformed to that of domain D by the mirror reflection with respect to the [010]--[001] plane. 
The helicity of the magnetic structure is reversed by the mirror reflection. 

\subsection{$\bm{D}$-vector in reciprocal space}
\begin{figure}[t]
\begin{center}
\includegraphics[width=8cm]{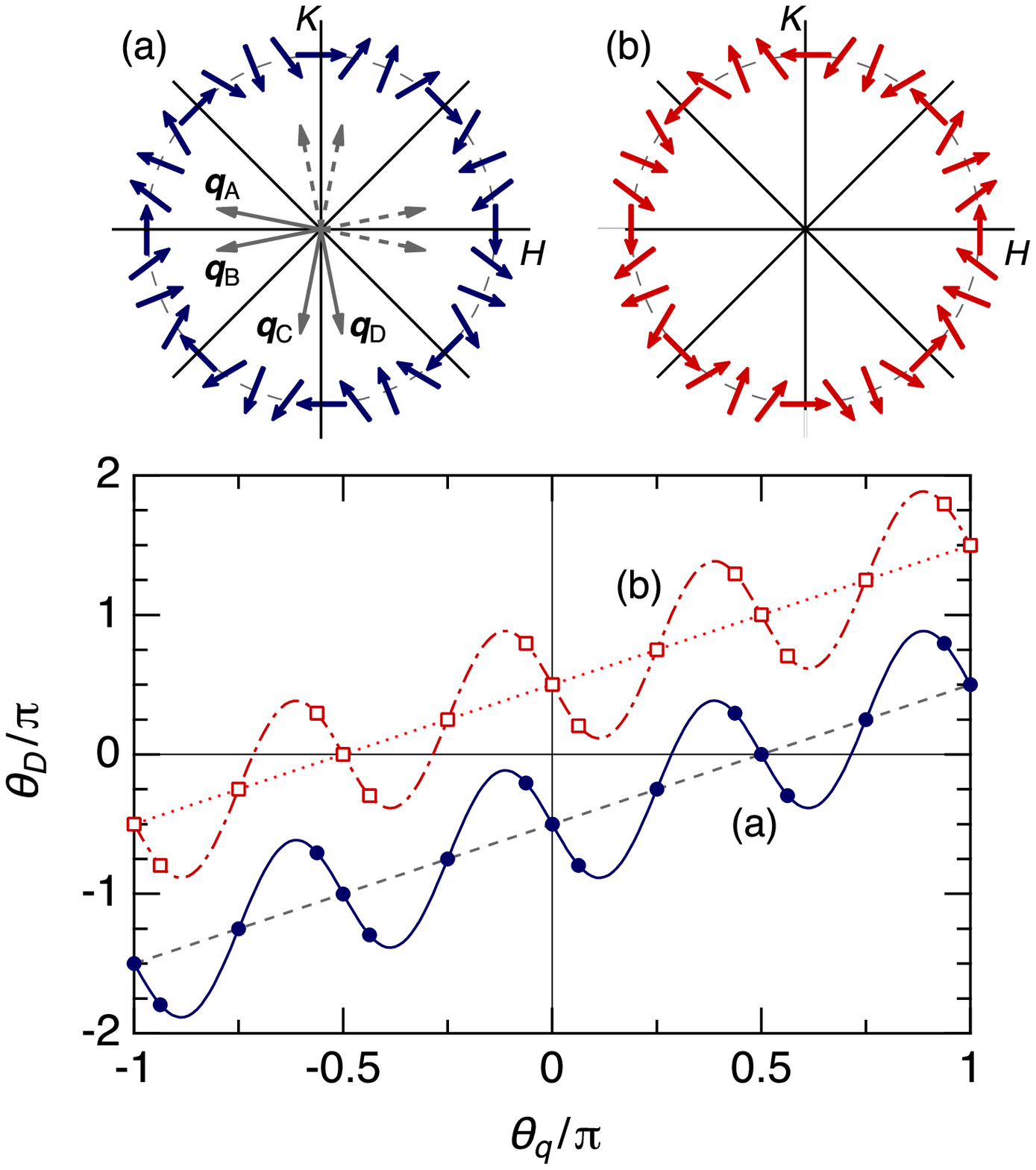}
\caption{$\bm{q}$ dependence of the $\bm{D}$ vector expected from the experimental $\bm{m}(\bm{q})$ vectors obtained at 0 T and the $C_{4v}$ symmetry of the crystal. 
}
\label{figS:DvecQ}
\end{center}
\end{figure}

In Fig.~2C of the main text, we presented a schematic distribution of the $\bm{D}$ vector in the reciprocal space. 
The $\bm{q}$-dependent $\bm{D}$ vector in the space group $I4mm$ lies in the $(HK0)$ plane~\cite{Yambe22}. 
When $\bm{q}$ is parallel to the mirror plane, the $\bm{D}$ vector is perpendicular to the mirror plane. 
This has been confirmed experimentally in the cycloidal order of EuIrGe$_3$~\cite{Kurauchi23}. 
When $\bm{q}$ is away from the mirror plane, such constraint is removed and $D_x(\bm{q})$ and $D_y(\bm{q})$ are independent. 
We are interested in this distribution of the $\bm{D}$ vector in the $HK$-plane. 
With respect to the eight $\bm{q}$ vectors, corresponding to the four helical domains at zero field, including $-\bm{q}$, we may assume that the $\bm{D}$ vector is perpendicular to the helical plane. 
We plot these eight points as experimental data in Fig.~\ref{figS:DvecQ}, where the directions of $\bm{q}$ and $\bm{D}$ are expressed by  
$\bm{q}=(\cos \theta_q, \sin \theta_q, 0)$ and $\bm{D}=(\cos \theta_D, \sin \theta_D, 0)$, respectively. 
Additional eight points for the $\bm{q}$ vectors on the mirror plane are also plotted so that the relation $\theta_D = -\pi/2+\theta_q$ holds ($\bm{D} \perp $ mirror plane).  This is the case (a) in Fig.~\ref{figS:DvecQ}. 

Next, the data points were fit with a simple sine wave as shown by the solid line. 
The distribution of the $\bm{D}$ vector in the $HK$-plane shown in Fig.~2C of the main text, corresponding to the case (a) here, was obtained in this way. 
Although some assumptions were used, the symmetry relation of the $\bm{D}$ vector in the $HK$-plane is visualized. 
The fourfold symmetry and the mirror-reflection symmetry are both satisfied. 
Note that we do not refer to the absolute direction of the $\bm{D}$ vector, i.e, which of (a) or (b) is the case. 
The experimental determination of the $\bm{D}$ vector is not in the scope of this work. 

\newpage
\subsection{Triple-$\bm{q}$ structure in phase II at 2.4 T}
\begin{figure}
\begin{center}
\includegraphics[width=8.5cm]{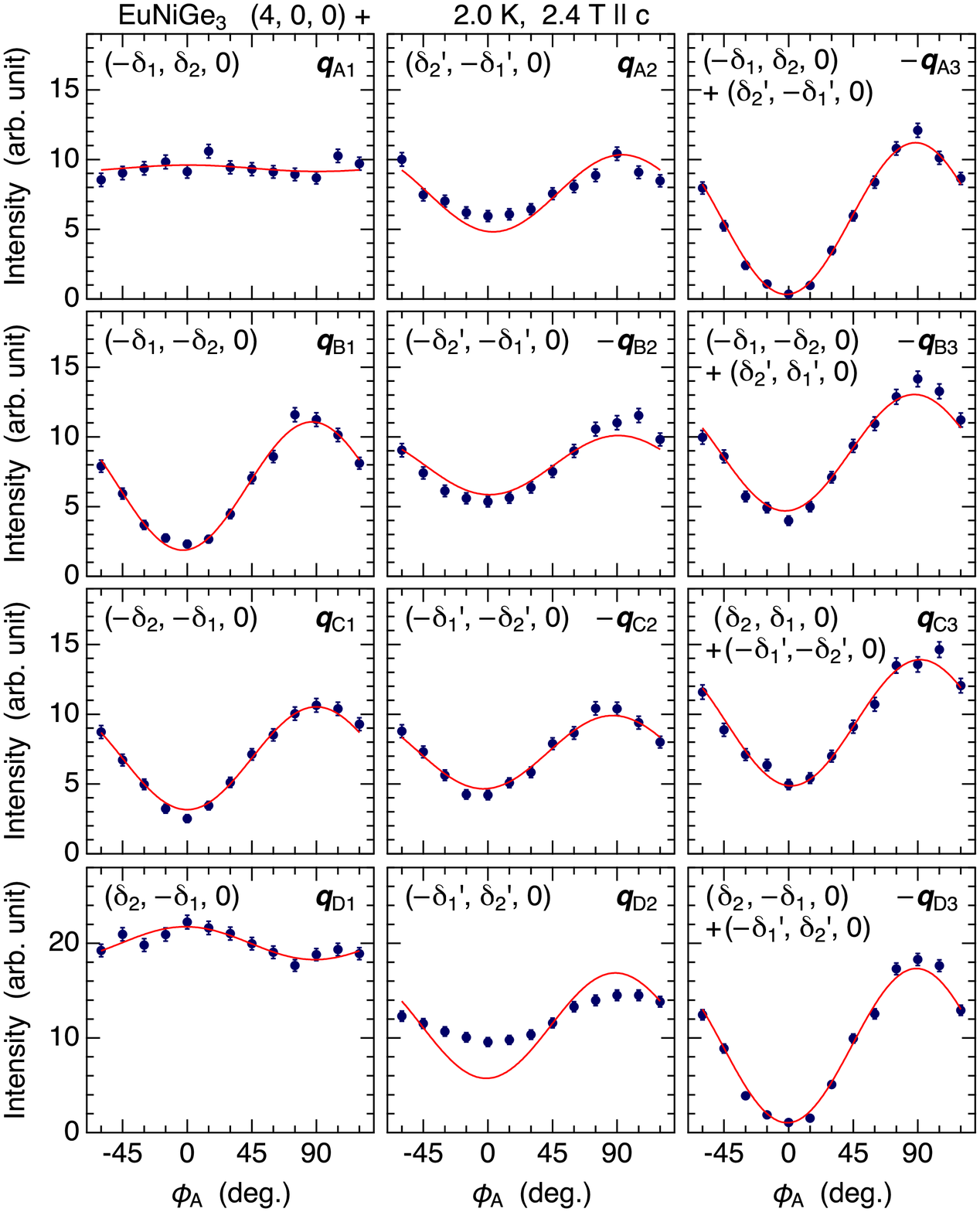}
\caption{
Analyzer angle ($\phi_{\text{A}}$) dependences of the magnetic Bragg-peak intensities around the (4, 0, 0) fundamental reflection at 2.4 T in phase II. 
The x-ray energy is 7.612 keV at resonance. 
The solid lines are the calculated intensity curves for the Fourier components summarized in Table~\ref{tbl:mq24kG}. 
}
\label{figS:POLscans24kG}
\end{center}
\end{figure}

\begin{figure}[t]
\begin{center}
\includegraphics[width=8.5cm]{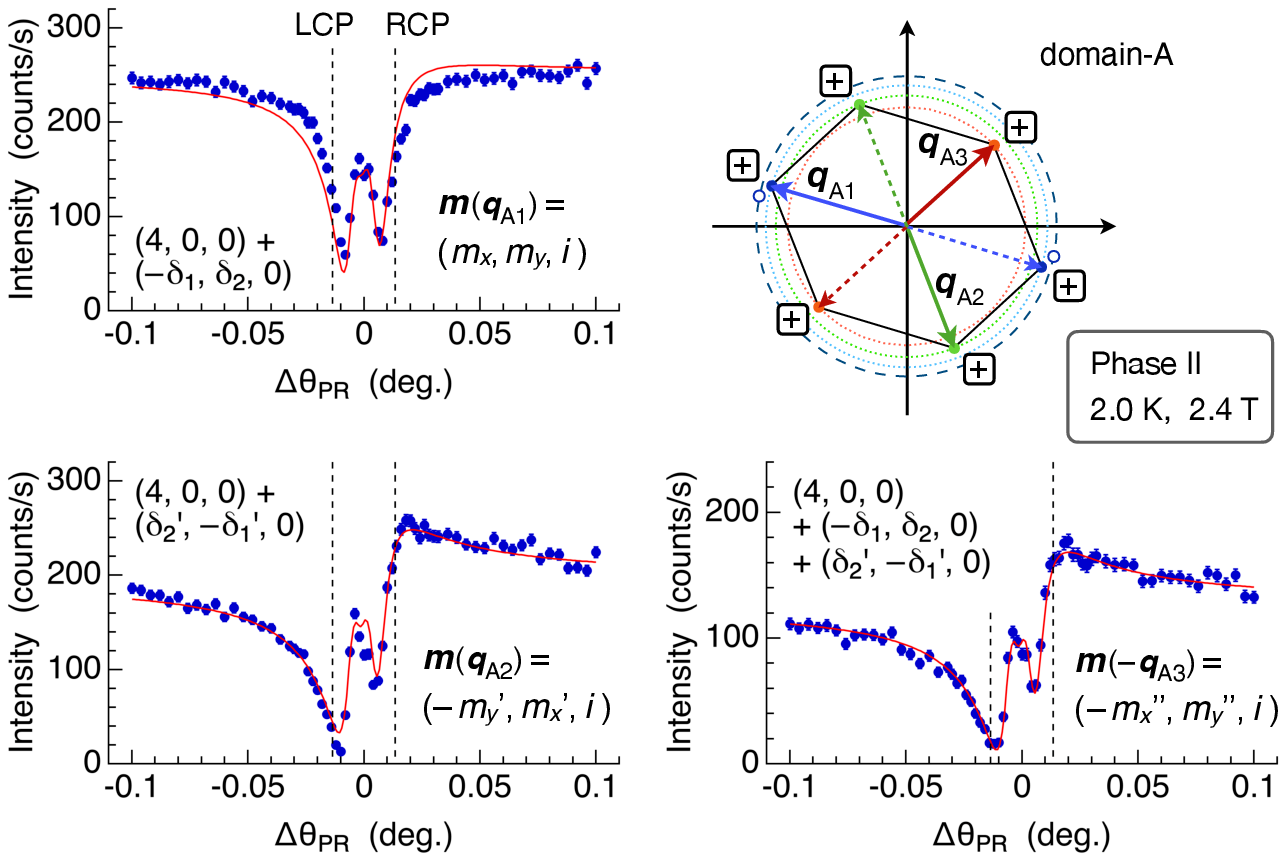}
\caption{$\Delta\theta_{\text{PR}}$ dependences of the magnetic Bragg-peak intensities at three $\bm{q}$ vectors for the domain-A in phase II at 2.4 T. The x-ray energy is 7.612 keV at resonance. The background has been subtracted.  The solid lines are the calculated intensity curves for the Fourier components summarized in Table~\ref{tbl:mq24kG}. 
}
\label{figS:PRth24kG_A}
\end{center}
\end{figure}

\begin{figure}[t]
\begin{center}
\includegraphics[width=8.5cm]{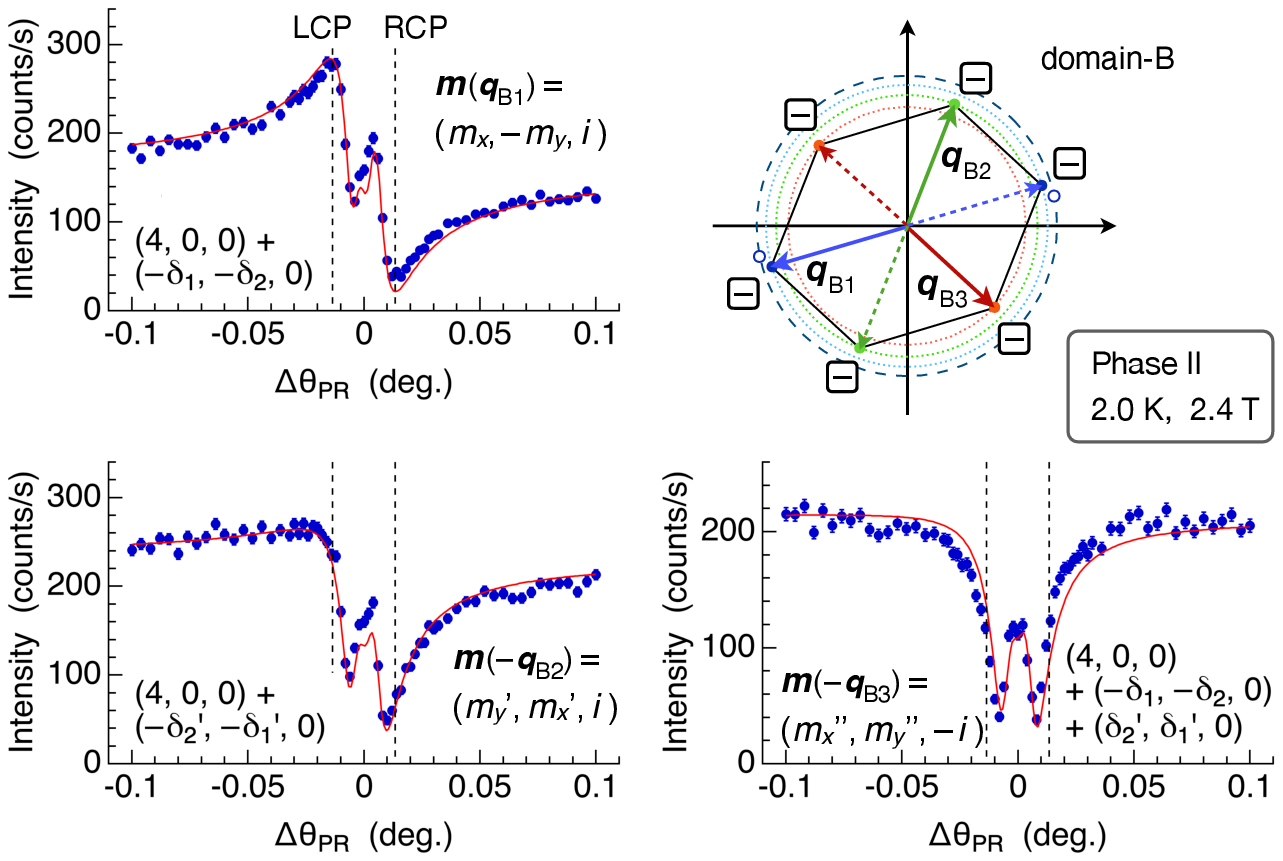}
\caption{$\Delta\theta_{\text{PR}}$ dependences of the magnetic Bragg-peak intensities at three $\bm{q}$ vectors for the domain-B in phase II at 2.4 T. The x-ray energy is 7.612 keV at resonance. The background has been subtracted. The solid lines are the calculated intensity curves for the Fourier components summarized in Table~\ref{tbl:mq24kG}. 
}
\label{figS:PRth24kG_B}
\end{center}
\end{figure}

\begin{figure}[t]
\begin{center}
\includegraphics[width=8.5cm]{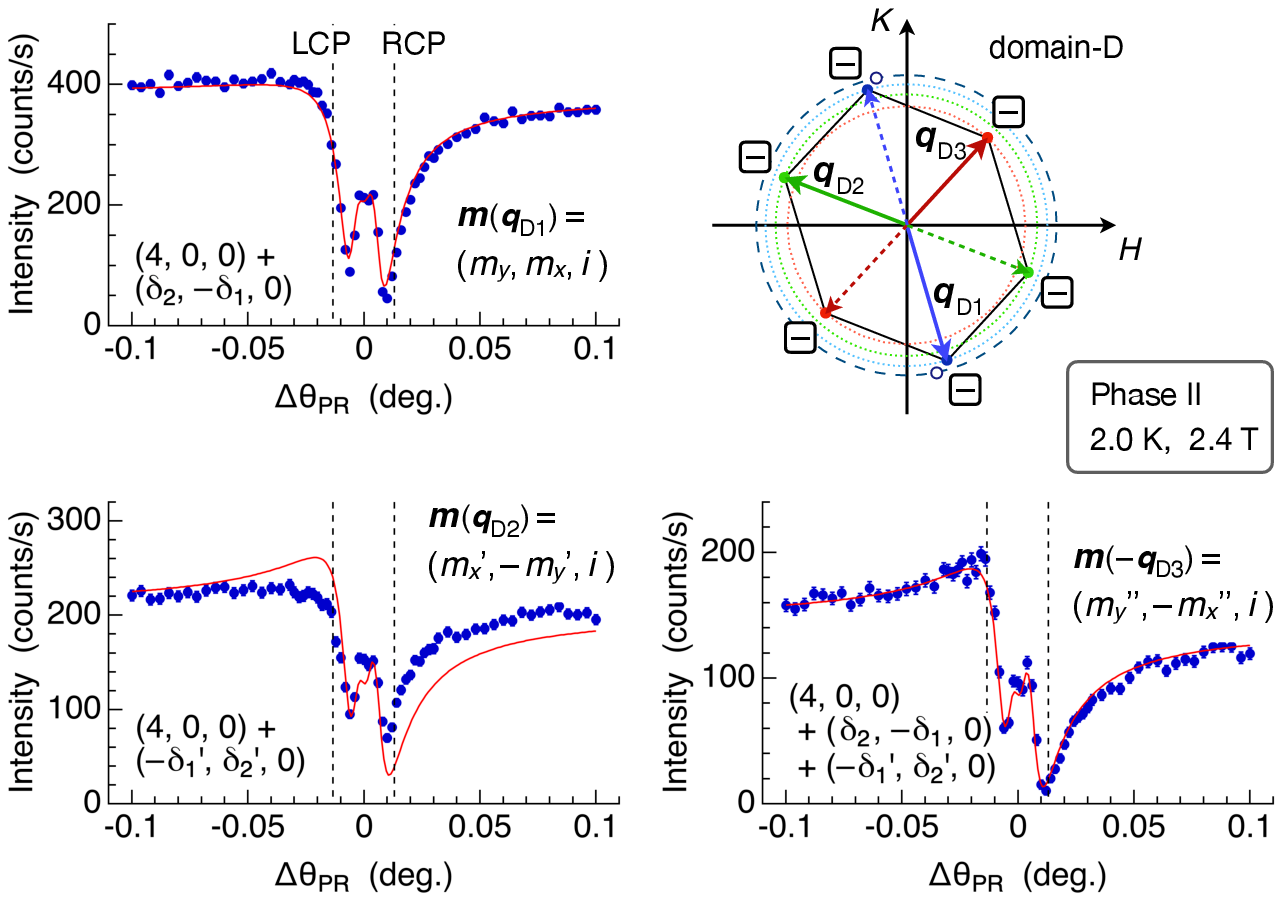}
\caption{$\Delta\theta_{\text{PR}}$ dependences of the magnetic Bragg-peak intensities at three $\bm{q}$ vectors for the domain-D in phase II at 2.4 T. The x-ray energy is 7.612 keV at resonance. The background has been subtracted. The solid lines are the calculated intensity curves for the Fourier components summarized in Table~\ref{tbl:mq24kG}. }
\label{figS:PRth24kG_D}
\end{center}
\end{figure}

The Fourier components of the triple-$\bm{q}$ magnetic structure in phase II at 2.4 T were obtained by the same method by combining the  $\phi_{\text{A}}$-scan and the $\Delta\theta_{\text{PR}}$-scan data sets. 
The results of linear polarization analysis ($\phi_{\text{A}}$ scan) for all the Bragg peaks in phase II, at three $\bm{q}$ vectors for each of the four domains, are summarized in Fig.~\ref{figS:POLscans24kG}. 
In Figs.~\ref{figS:PRth24kG_A}, \ref{figS:PRth24kG_B}, and \ref{figS:PRth24kG_D}, the results of the $\Delta\theta_{\text{PR}}$ scan for domains A, B, and D, respectively, are summarized in the same manner as for the domain C in Fig.~3 of the main text. 
The Fourier components of $\bm{m}(\bm{q}_1)$, $\bm{m}^{\prime}(\bm{q}_2)$, and $\bm{m}^{\prime\prime}(\bm{q}_3)$ are summarized in Table~\ref{tbl:mq24kG}, where the $x$ and $y$ components are determined to satisfy the $C_{4v}$ symmetry and simultaneously explain the data sets as much as possible. 
The $z$ ($c$-axis) components of $\bm{m}(\bm{q})$, which are expressed by an imaginary number, determines the sense of rotation of the helimagnetic Fourier components. This is sensitively reflected in the asymmetry in the $\Delta\theta_{\text{PR}}$-scan data. 
The helicities determined experimentally, which are also summarized in Table~\ref{tbl:mq24kG}, clearly show that each domain of the triple-$\bm{q}$ magnetic structure has its own unified helicity that is primarily determined by the parental helicity of $\bm{m}(\bm{q}_1)$.

\begin{table}
\caption{The $\bm{q}$ vectors and the Fourier components of the triple-$\bm{q}$ helical magnetic structure at 2.4 T in phase II. 
The $\bm{q}$ vectors are chosen so that the relation $\bm{q}_1 + \bm{q}_2 + \bm{q}_3 = 0$ is satisfied.
The parameters are $(\delta_1, \delta_2)=(0.237, 0.072)$, $(\delta_1^{\prime}, \delta_2^{\prime})=(0.215, 0.083)$, 
$(m_x, m_y)=(0.41, 1.03)$, $(m_x^{\prime}, m_y^{\prime})=(0.90, 0.09)$, and $(m_x^{\prime\prime}, m_y^{\prime\prime})=(0.28, 0.56)$. 
The angle between $\bm{q}$ and $\bm{m}(\bm{q})$ is $85 \pm 5^{\circ}$ for $\bm{q}_1$, $65 \pm 5^{\circ}$ for $\bm{q}_2$, and $76 \pm 5^{\circ}$ for $\bm{q}_3$.  
The ratio between the $z$ ($c$-axis) and the $xy$ ($ab$-plane) components of $\bm{m}(\bm{q})$, i.e., $1/\sqrt{m_x^{\;2} + m_y^{\;2}}$, is  $0.9 \pm 0.1$ for $\bm{m}(\bm{q}_1)$, $1.1 \pm 0.1$ for $\bm{m}^{\prime}(\bm{q}_2)$, and $1.6 \pm 0.1$ for $\bm{m}^{\prime\prime}(\bm{q}_3)$. }
\label{tbl:mq24kG}
\begin{tabular}{cllc}
\hline
domain & \hspace{10mm} $\bm{q}$ & \hspace{3mm} $\bm{m}(\bm{q})$ & helicity \\
\hline
A & $\bm{q}_{\text{A1}}$=$(-\delta_1, \delta_2, 0)$ & $( m_x, m_y, i)$  & $+$ \\
  & $\bm{q}_{\text{A2}}$=$(\delta_2^{\prime}, -\delta_1^{\prime}, 0)$ & $( -m_x^{\prime}, m_y^{\prime}, i)$  & $+$ \\
  & $\bm{q}_{\text{A3}}$=$(\delta_1 - \delta_2^{\prime}, -\delta_2 + \delta_1^{\prime}, 0)$ & $( m_x^{\prime\prime}, -m_y^{\prime\prime}, i)$  & $+$ \\
B & $\bm{q}_{\text{B1}}$=$(-\delta_1, -\delta_2, 0)$ & $( m_x, -m_y, i)$  & $-$ \\
  & $\bm{q}_{\text{B2}}$=$(\delta_2^{\prime}, \delta_1^{\prime}, 0)$ & $( -m_x^{\prime}, -m_y^{\prime}, i)$  & $-$ \\
  & $\bm{q}_{\text{B3}}$=$(\delta_1 - \delta_2^{\prime}, \delta_2 - \delta_1^{\prime}, 0)$ & $( m_x^{\prime\prime}, m_y^{\prime\prime}, i)$  & $-$ \\
C & $\bm{q}_{\text{C1}}$=$(-\delta_2, -\delta_1, 0)$ & $( -m_y, m_x, i)$  & $+$ \\
  & $\bm{q}_{\text{C2}}$=$(\delta_1^{\prime}, \delta_2^{\prime}, 0)$ & $( -m_y^{\prime}, -m_x^{\prime}, i)$  & $+$ \\
  & $\bm{q}_{\text{C3}}$=$(\delta_2 - \delta_1^{\prime}, \delta_1 - \delta_2^{\prime}, 0)$ & $( m_y^{\prime\prime}, m_x^{\prime\prime}, i)$   & $+$ \\
D & $\bm{q}_{\text{D1}}$=$(\delta_2, -\delta_1, 0)$ & $( m_y, m_x, i)$  & $-$ \\
  & $\bm{q}_{\text{D2}}$=$(-\delta_1^{\prime}, \delta_2^{\prime}, 0)$ & $( m_y^{\prime}, -m_x^{\prime}, i)$  & $-$ \\
  & $\bm{q}_{\text{D3}}$=$(-\delta_2 + \delta_1^{\prime}, \delta_1 - \delta_2^{\prime}, 0)$ & $( -m_y^{\prime\prime}, m_x^{\prime\prime}, i)$  & $-$ \\
\hline
\end{tabular}
\end{table}

\newpage
\subsection{Higher-order peak in phase II at 2.4 T}
\begin{figure}[t]
\vspace{10mm}
\begin{center}
\includegraphics[width=8.5cm]{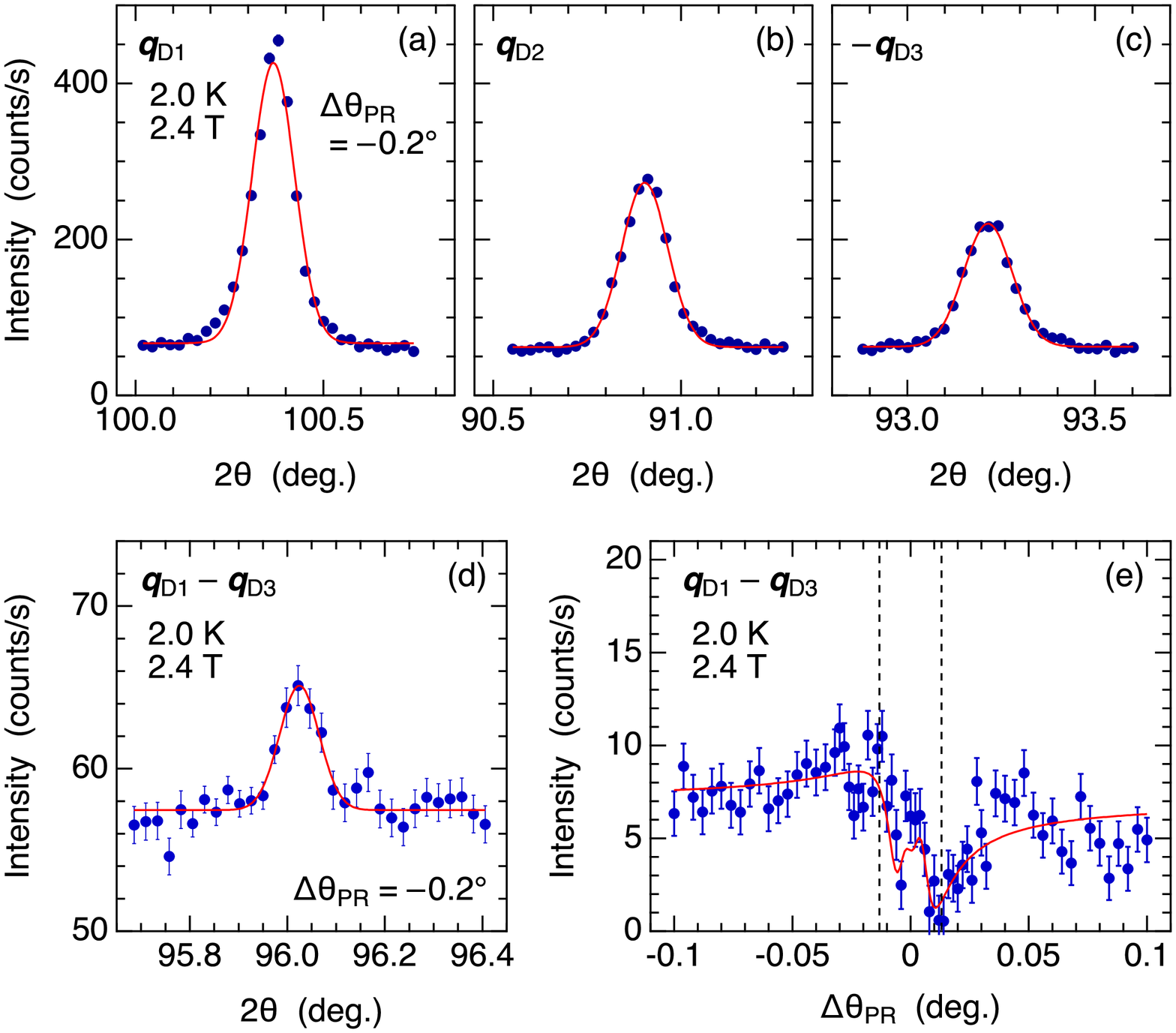}
\caption{(a, b, c) Peak profile of the $\omega$-$2\theta$ scan for the magnetic Bragg peak at $\bm{q}_{\text{D1}}$, $\bm{q}_{\text{D2}}$, and $-\bm{q}_{\text{D3}}$, respectively, in phase II at 2.4 T. The phase retarder is set at $\Delta\theta_{\text{PR}}=-0.2^{\circ}$.  
(d) Peak profile of the $\omega$-$2\theta$ scan for the higher-order peak at $\bm{q}_{\text{D1}}-\bm{q}_{\text{D3}}=(2\delta_2-\delta_1^{\prime},-2\delta_1+\delta_2^{\prime}, 0)$. 
(e) $\Delta\theta_{\text{PR}}$ dependence of the intensity of the higher-order peak at $\bm{q}_{\text{D1}}-\bm{q}_{\text{D3}}$. 
The background has been subtracted. 
The solid line is a calculated intensity curve for the Fourier component $\bm{m}(\bm{q}_{\text{D1}}-\bm{q}_{\text{D3}})=(1.00, -0.07, i)$ with a helicity ($-$). 
The x-ray energy is 7.612 keV at resonance.
}
\label{figS:PRth24kG_D13}
\end{center}
\end{figure}

The peak profile of the magnetic Bragg peak at $\bm{q}_{\text{D1}}$, $\bm{q}_{\text{D2}}$, and $-\bm{q}_{\text{D3}}$ for the domain-D in phase II is shown in Fig.~\ref{figS:PRth24kG_D13}(a), (b), and (c), respectively. The phase retarder is used and the polarization analysis is not performed (Fig. \ref{figS:config}). 
The intensity of the $\bm{q}_{\text{D1}}$ peak is stronger by a factor of $\sim 2$ than that of the $\bm{q}_{\text{D3}}$ peak. 
This is because the $\bm{q}_{\text{D1}}$ intensity consists of both $\pi$-$\pi'$ and $\pi$-$\sigma'$ scatterings, whereas the $\bm{q}_{\text{D3}}$ intensity consists mostly of $\pi$-$\pi'$ scattering only as shown in Fig.~\ref{figS:POLscans24kG}. 
This is due to the geometrical reason. The intensities of the $\pi$-$\pi'$ scattering for the three peaks are almost equal. 

The higher-order peak at $\bm{q}_{\text{D1}}-\bm{q}_{\text{D3}}=(2\delta_2-\delta_1^{\prime},-2\delta_1+\delta_2^{\prime}, 0)$ is shown in Fig.~\ref{figS:PRth24kG_D13}(d). The intensity is $\sim 3$\% of the average of the $\bm{q}_{\text{D1}}$ and $\bm{q}_{\text{D3}}$ intensities. 
The result of the $\Delta\theta_{\text{PR}}$ scan for this higher-order peak is shown in Fig.~\ref{figS:PRth24kG_D13}(e). 
Note that $\bm{q}_{\text{D1}}-\bm{q}_{\text{D3}}$ is in the \textit{positive} helicity region as for $\bm{q}_{\text{C}}$ at zero field. 
However, the asymmetry of the $\Delta\theta_{\text{PR}}$ dependence for the $\bm{q}_{\text{D1}}-\bm{q}_{\text{D3}}$ peak is clearly opposite to that for the $\bm{q}_{\text{C}}$ peak at zero field (see Fig. 2B of the main text), indicating that the helicity of this Fourier component is ($-$). 
Since all three primary components of $\bm{q}_{\text{D1}}$, $\bm{q}_{\text{D2}}$, and $\bm{q}_{\text{D3}}$ have \textit{negative} helicity, this is a natural result. The unified helicity extends to higher-order Fourier components. 
Although it is difficult to determine precisely the Fourier component of the higher-order peak due to the relatively poor statistics, the intensity calculation by assuming $\bm{m}(\bm{q}_{\text{D1}}-\bm{q}_{\text{D3}})$ is approximately around the average of $\bm{m}(\bm{q}_{\text{D1}}$ and $\bm{m}(\bm{q}_{\text{D3}}$, i.e., $\bm{m}(\bm{q}_{\text{D1}}-\bm{q}_{\text{D3}}) \sim (1.00, -0.07, i)$, well explains the data. 

\end{document}